\documentclass[journal=jctcce,manuscript=article,layout=traditional]{achemso}

\usepackage{chemformula} 
\usepackage[T1]{fontenc} 
\usepackage{braket}
\usepackage{mathtools}
\usepackage{float}
\usepackage{longtable}
\usepackage{setspace}

\definecolor{Gray}{gray}{0.9}

\usepackage{color, colortbl}

\def\br{{\mathbf{r}}}

\usepackage{bm}
\def\bPsi{{\bm{\Psi}}}
\def\bPhi{{\bm{\Phi}}}
\def\bV{{\bm{V}}}
\def\bG{{\bm{G}}}
\def\bP{{\bm{P}}}
\def\bX{{\bm{X}}}
\def\bY{{\bm{Y}}}

\def\cE{{\mathcal E}}
\def\cL{{\mathcal L}}
\def\cF{{\mathcal F}}
\def\cU{{\mathcal U}}
\def\cT{{\mathcal T}}

\author{Aleksei V. Ivanov}
\affiliation[University Iceland]
{Science Institute and Faculty of Physical Sciences, University of Iceland, VR-III, 107
Reykjav\'{i}k, Iceland}

\author{Gianluca Levi}
\affiliation[University Iceland]
{Science Institute and Faculty of Physical Sciences, University of Iceland, VR-III, 107
Reykjav\'{i}k, Iceland}

\author{Elvar \"O. J\'{o}nsson}
\affiliation[University Iceland]
{Science Institute and Faculty of Physical Sciences, University of Iceland, VR-III, 107
Reykjav\'{i}k, Iceland}

\author{Hannes J\'{o}nsson}
\affiliation[University Iceland]
{Science Institute and Faculty of Physical Sciences, University of Iceland, VR-III, 107
Reykjav\'{i}k, Iceland}
\email{hj@hi.is}

\title[Direct Orbital Optimization of Excited States]
{Method for Calculating Excited Electronic States Using Density Functionals and Direct Orbital Optimization with Real Space Grid or Plane Wave Basis Set}


\begin{document}

\begin{tocentry}
    \includegraphics[width=1.0\textwidth]{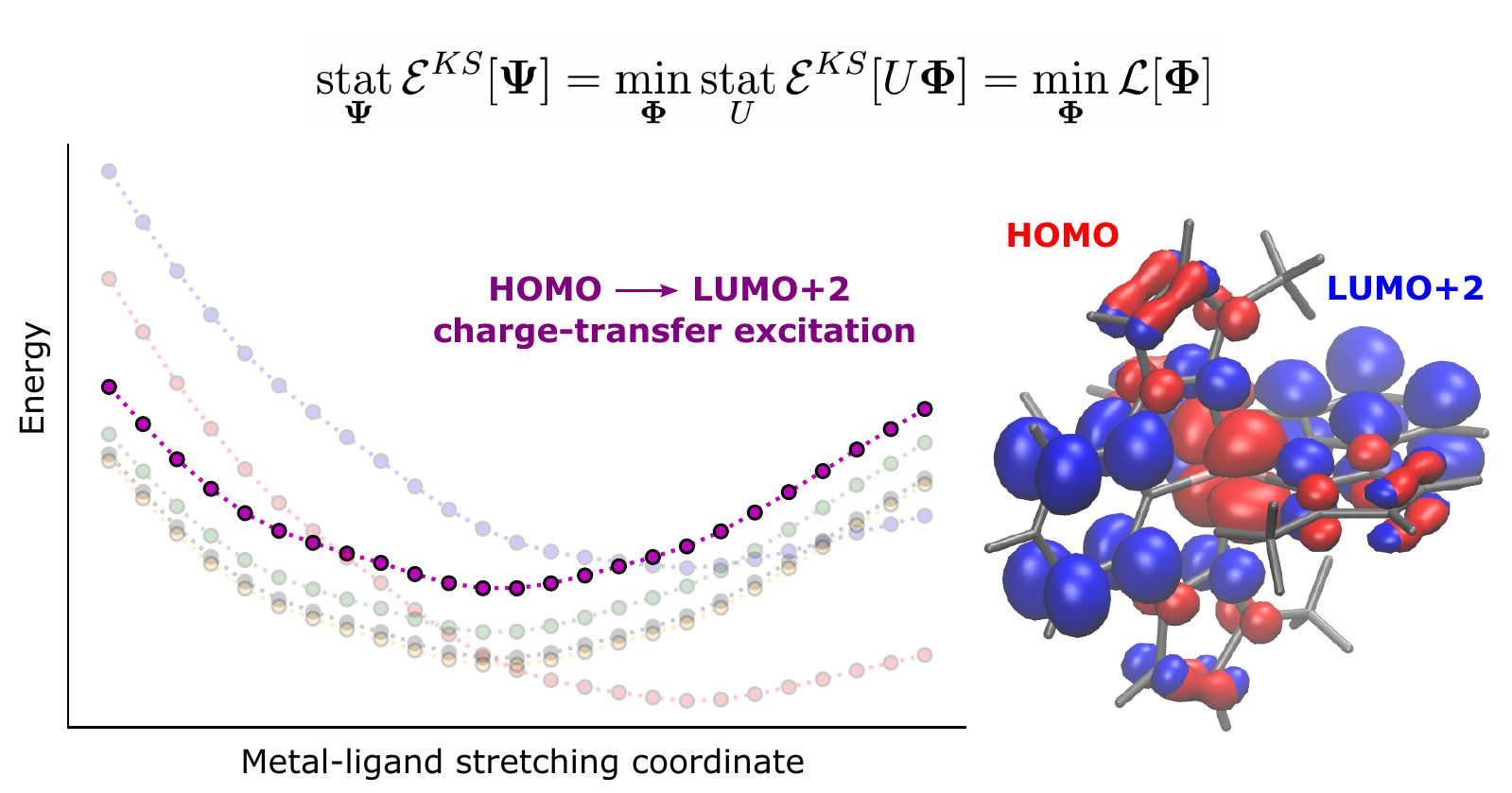}
\end{tocentry}

\begin{abstract}
A direct orbital optimization method is presented for density functional calculations of excited electronic states using either 
a real space grid or a plane wave basis set. 
The method is variational, provides atomic forces in the excited states, and can be 
applied to Kohn-Sham (KS) functionals as well as orbital-density dependent functionals (ODD) 
including explicit self-interaction correction.
The implementation for KS functionals involves two nested loops: 
(1) An inner loop for finding a stationary point in a subspace spanned by the occupied and a few virtual orbitals corresponding to the excited state; 
(2) an outer loop for minimizing the energy in a tangential direction in the space of the orbitals.
For ODD functionals, a third loop is used to find the unitary transformation that minimizes the energy functional among occupied orbitals only.
Combined with the maximum overlap method, the algorithm
converges in challenging cases where conventional self-consistent field algorithms tend to fail. 
The benchmark tests presented include two charge-transfer excitations in 
nitrobenzene and an excitation of CO to degenerate $\pi^\ast$ orbitals where the importance of complex orbitals is illustrated. 
An application of the method to
several metal-to-ligand charge-transfer and metal-centred excited states of an Fe$^{\rm II}$ photosensitizer complex is described 
and the results compared to reported experimental estimates.
The method is also used to study 
the effect of Perdew-Zunger self-interaction correction on valence and Rydberg excited states of several molecules, 
both singlet and triplet states, and the performance compared to semilocal and hybrid functionals.
\end{abstract}

\section{Introduction}

Density functional theory (DFT) is commonly used in computational studies of molecules and materials as it can in many cases 
give reasonable accuracy without too much computational effort. Calculations of ground electronic states can be performed 
even by non-experts thanks to well-established algorithms and software implementations. This does not, however, apply to 
calculations of excited electronic states, although such states are of great importance in many rapidly developing fields 
such as ultrafast spectroscopy, solar energy conversion and photocatalysis. 
The most commonly used excited-state extension of DFT is time-dependent density functional theory (TDDFT)~\cite{Runge1984,Casida1995, Dreuw2005}. In practical implementations, TDDFT calculations are carried out using 
some ground-state density functional, linear-response theory and an adiabatic approximation that neglects the time dependence of
the exchange-correlation (XC) kernel. Within these approximations, TDDFT typically provides a fairly good description of
low-lying valence excitations~\cite{Dreuw2005}, 
but often fails to describe higher excitations~\cite{Levine2006, Maitra2004, Tozer2000}, long-range charge-transfer states~\cite{Dreuw2004a} and conical intersections 
between ground and excited states~\cite{Levine2006}.

Alternative approaches with wide applicability and similar computational effort can be based on
time-independent DFT. These include ensemble DFT~\cite{Oliveira1990, Yang2017, Deur2017}, excited-state DFT (eDFT)~\cite{Hellman2004,Gavnholt2008,Cheng2008,Gilbert2008,Kowalczyk2011,Levi2020,Hait2020,CarterFenk2020} 
(also sometimes referred to as $\Delta$ self-consistent field, $\Delta$SCF), as well as 
constrained DFT~\cite{Ramos2018,Roychoudhury2020,Karpinski2020}, orthogonality constrained~\cite{Baruah2009,Baruah2012,Evangelista2013},
and constricted DFT approaches~\cite{Ziegler2012,Park2016}. 
There are also methodologies where excited-state properties are obtained from ground-state calculations only, as for example in electron-hole self-interaction corrected calculations~\cite{Harrison1983,Pederson1988} or the quasi-particle energy DFT (QE-DFT)~\citep{Mei2019}.
In eDFT, excited states are found as stationary states of the energy expressed as a density functional. This approach is often considered unjustified because DFT is formulated as a ground state theory, where it is based on a one-to-one map between the density and the external potential. Even if this correspondence does not exist in the case of excited states and arbitrary external potentials~\cite{Gaudoin2004}, generalizations of DFT to excited states are still possible~\cite{Levy1999,Gorling1999,Ayers2012,Ayers2015,Ayers2018}. The excited-state functional can be constructed as a bifunctional depending on both ground- and excited-state densities~\cite{Levy1999}. Moreover, for finite systems with a Coulomb external potential, which includes most systems of relevance, the Coulomb density uniquely defines the potential and different excited states cannot have the same Coulomb density~\cite{Ayers2012,Ayers2018}. Therefore, for finite Coulomb systems, the excited-state energy can be expressed as a unique functional of the excited-state density only. As in the case of the ground state, Kohn-Sham equations for excited states can also be introduced~\cite{Ayers2015}. Practical KS eDFT calculations use two approximations: (1) ground-state exchange-correlation functionals are employed, and (2) the Slater determinants corresponding to excited and ground states are in general not orthogonal. Regarding the second approximation, one can employ an orthogonalization procedure after the SCF calculations in order to obtain excited-state properties such as transition dipole moments~\cite{Bourne-Worster2021}. In the present work, we consider only finite Coulomb systems.

Since in eDFT an 
excited state is found as a solution of the Kohn-Sham (KS) equations for non-aufbau orbital occupation numbers, the computational robustness strongly depends on the algorithm used to solve the self-consistent field equations.
Commonly used methods in
ground-state calculations are based on 
some iterative eigensolver such as the Davidson algorithm~\cite{Davidson1975} or the
residual minimization method--direct inversion in the iterative subspace (RMM-DIIS)~\cite{Pulay1980,Pulay1982,Kresse1996prb}
enhanced in various ways to improve robustness and rate of convergence~\cite{Kresse1996,Garza2012}. 
However, these algorithms are not specifically designed for calculations of excited states. 
The maximum overlap method (MOM)~\cite{Gilbert2008} can be used to 
reduce the probability of convergence on the ground state in the iterative calculation, 
but convergence problems often occur~\cite{Hait2020}. 
The basic problem lies in the fact that excited-state calculations do not involve finding the global minimum of the energy as a 
function of the electronic degrees of freedom, but rather a more general stationary point on the high-dimensional electronic energy surface.
A method that can in general converge on a saddle point on the energy surface
rather than a minimum is required.
To find an N-th order saddle point, one needs to maximize the energy with respect to N degrees of freedom 
while minimizing with respect to all the others. 
The degrees of freedom along which the energy needs to be maximized are not known {\it a priori}. 
This makes a search for a saddle point significantly more difficult than a search for a minimum. 
Therefore, while calculations based on 
SCF-MOM-type algorithms can in principle converge on excited states, they are in practice not reliable.

Alternatively, direct optimization (DO) can be used to converge on 
solutions of the KS equations.
While this approach has mainly been used in energy minimization to find a ground-state solution~\citep{Gillan1989,Payne1992,Hutter1994,Marzari1997,Ismail-Beigi2000a,VanVoorhis2002,VandeVondele2003,Weber2008,Freysoldt2009}, it can be extended to calculations of excited states~\cite{Ye2017,Levi2020,Hait2020}.
 One possible approach is to formulate the problem as a search for a minimum of the norm of the gradient~\cite{Hait2020}. 
 But, it is then important to also ensure that the norm of the gradient is zero at the minimum
 and the evaluation of the gradient of this objective function
 increases the computational effort significantly.
 A different approach is based on direct optimization of the energy using a quasi-Newton method that can develop negative eigenvalues of the Hessian consistent with the type of saddle point searched for~\cite{Levi2020, Levi2020fd}. 
 A preconditioner is then needed to estimate the degrees of freedom for which the energy needs to be maximized 
 and thereby ensure convergence on the desired saddle point. When combined with MOM, this approach can perform better than conventional SCF-MOM algorithms~\cite{Levi2020,Levi2020fd}. 
 The number of degrees of freedom for which the optimization needs to be carried out is, however, an important consideration and so far the DO-MOM approach has only been formulated and implemented in the context of the linear combination of atomic orbitals (LCAO) basis set 
 where the number of degrees of freedom is relatively small.

 Real space grid (RSG) and plane wave (PW) basis sets have the advantage that the complete basis set limit can be reached by
 varying systematically a single parameter such as the mesh spacing or plane wave cutoff. 
 They are also more universal and can easily be applied to 
 diffuse states such as Rydberg and metallic states where
 a typical LCAO basis set needs to be supplemented by specially tailored diffuse basis functions. 
 This is demonstrated for Rydberg states of NH$_3$ in section 3 of this article.
 In calculations 
relevant to, for example, ultrafast experiments, the system may evolve through a series of localized and delocalized states,  
 making it challenging to design an LCAO basis set that is complete enough for all the relevant states. It is clearly more convenient to use an RSG or PW basis set in these cases.


In this article, a DO-MOM algorithm for calculating excited electronic states that can be used with both RSG and PW basis sets is presented. 
The method is based on  an auxiliary energy functional that has a minimum at a stationary solution of the Kohn-Sham functional. The auxiliary functional is constructed by introducing an inner loop, which finds a stationary point of the energy through a unitary transformation in a reduced space of occupied and a few virtual orbitals.
The calculations are variational\footnote{The term variational is used here in the sense that the calculated energy of an excited state is stationary with respect to any infinitesimal change in the orbitals satisfying orthonormality constraints. If two or more solutions of the KS equations correspond to the same excited state then the lowest energy excited state is not necessarily the most appropriate. It may, for example, provide a density that lacks the symmetry of the molecule, as discussed in Sec.~\ref{sec: numtest}} and, since the Hellmann-Feynman theorem is satisfied not only at the minimum but also at 
any stationary point on the electronic energy surface, they can be used to evaluate the
atomic forces, thereby providing a powerful tool for exploring excited-state potential energy surfaces (PESs) 
in, for example, simulations of the dynamics or minimum energy path calculations~\cite{Asgeirsson2020}.
The DO-MOM method is tested and its performances compared with that of SCF-MOM with either Davidson or RMM-DIIS methods combined with
Pulay density mixing~\cite{Pulay1980}.
The tests include 
charge-transfer excitations in nitrobenzene that are known to be challenging cases for conventional algorithms. 
While the SCF-MOM algorithms show erratic behavior, likely because of the presence of several orbitals with 
similar energy, 
the DO-MOM calculation converges in a robust way.
Another test involves calculations of an electronic excitation of the CO molecule to degenerate states. 
Again, SCF-MOM shows erratic behavior while the DO-MOM calculation converges smoothly.
There, the advantage of using complex orbitals instead of real orbitals is, furthermore, demonstrated.  

Two applications of the DO-MOM method are presented. 
The first one involves calculations of metal-to-ligand charge-transfer (MLCT) and metal-centered (MC) excitations 
of the [Fe(bmip)$_2$]$^{2+}$ (bmip=2,6-bis(3-methyl-imidazole-1-ylidine)-pyridine) complex, 
a prototype of a class of Fe-based photosensitizers~\cite{Liu2013,Harlang2015,Lindh2020}.
This complex has been studied experimentally using X-ray emission and scattering with femtosecond resolution~\cite{Kunnus2020} and 
theoretically using TDDFT~\cite{Papai2016,Papai2019a}.
In order to optimize its performance as a photosensitizer, an understanding of the transitions between the various states
and the way they are affected by the ligands is needed. 
An initial excitation to a singlet MLCT state is believed to be followed in part by a relaxation to a lower energy triplet MLCT state 
while another part
decays to a triplet MC state where it generates a vibrational wavepacket along a metal-ligand bond stretching coordinate~\cite{Papai2016,Papai2019a}. 
This branching  
occurs on ultrafast time scale, on the order of 100 fs, and influences the performance of the complex as photosensitizer. 
Dynamics of the molecule in the lowest-lying, dark singlet MLCT state has been simulated using energy surfaces calculated with TDDFT
but a higher energy, bright singlet MLCT state is likely populated in the experiments.~\cite{Papai2016,Papai2019a}
As a result, direct comparison with the ultrafast 
branching observed in the experiment could not be made.
Here, the DO-MOM method is used to calculate six excited states that are close in energy, 
including the bright singlet MLCT state along the metal-ligand bond stretching coordinate that is believed to be activated during the photoinduced dynamics. 
The calculated excitation energy agrees well with the experimentally observed value
and, furthermore, an estimate of the vibrational period in the lowest triplet MC state 
is also found to be in good agreement with experimental observations.
This demonstrates that the DO-MOM method could, in future work, be used in dynamics simulations 
to help interpret the experiments on this and similar photosensitizer complexes.

To illustrate the applicability of the present DO-MOM method to orbital-density dependent (ODD) functionals,
the effect of Perdew-Zunger self-interaction correction (PZ-SIC)~\cite{Perdew1981} 
on excitation energies is investigated. Semilocal approximations of the XC energy possess a spurious self-interaction error due to their inability
to cancel the non-local self-interaction error in the classical Coulomb energy. Perdew and Zunger proposed a correction~\cite{Perdew1981} in which the self-interaction error is estimated on each molecular spin-orbital separately and subtracted from the total energy. The resulting PZ functional is ODD and is not 
invariant to rotations among equally occupied orbitals. The DO-MOM method generalizes variational calculations of excited states to non-unitary invariant functionals. This permits an assessment
of the effect of the self-interaction error inherent in semilocal functionals on excited states. Here, the excitation energies of 13 transitions to singlet and triplet excited states in 9 molecules are calculated with the semilocal PBE~\cite{Perdew1996}, the self-interaction corrected PBE-SIC/2~\cite{Klupfel2012mol} and PBE-SIC~\cite{Perdew1981}, and the hybrid PBE0~\cite{PBE0} and PBE50 (50\% of exact exchange) functionals.
The results are compared with theoretical best estimates as well as experimental values.
Furthermore, O-H bond stretching curves in the water molecule and water dimer are calculated.
It is found that PZ-SIC improves the shape of the curves, producing a local minimum analogous to what has been found in high level wave function calculations while the uncorrected functional does not.

The article is organized as follows. 
In section~\ref{sec: methodology}, the DO-MOM algorithm for variational density functional calculations of excited states is presented. Section~\ref{sec: numtest} shows the results of the numerical tests. 
In sections~\ref{sec: application} and~\ref{sec: PZ-SIC}, the applications of the DO-MOM method to excited-state energy curves of the Fe$^{\rm II}$ complex
and the effect of PZ-SIC on excited states of molecules are presented, respectively.
Finally, a discussion and conclusions are presented in section~\ref{sec: discandconc}.


\section{Methodology}\label{sec: methodology}

In generalized KS-DFT, the energy of an electronic system is given by
\begin{equation}\label{eq: KSenergy}
\cE[\bPsi] = \cT_{s} + \int d^{3} \br \rho(\br) v_{ext}(\br) + \cU[\rho] + \cE_{xc},
\end{equation}
where $\cT_{s}$ is the kinetic energy of a system of non-interacting electrons that have the same density $\rho(\br)$ as the interacting electron system
\begin{equation}\label{eq: kinen}
\cT_s = -\frac{1}{2}\sum_{i=1}^{M} \sum_{\sigma=0,1} f_{i\sigma} \Braket{\psi_{i\sigma}|\nabla^{2}|\psi_{i\sigma}}
\end{equation}
and
\begin{equation}\label{eq: density}
\sum_{i=1}^{M}\sum_{\sigma=0,1} f_{i\sigma} \braket{\br|\psi_{i\sigma}} \braket{\psi_{i\sigma}| \br } = \rho(\br).
\end{equation}
with occupation numbers $\{f_{i\sigma}\}$. 
The occupation numbers can be chosen to be non-aufbau in order to represent an excited state. 
$M$ is the number of bands (orbitals) in the calculations and $\sigma$ is the spin index.
$v_{ext}(\br)$ is the external potential and 
 $\cU$ is the classical Coulomb energy
\begin{equation}
\cU[\rho] = \frac{1}{2}\iint d^3{\bf r}\,d^3{\bf r'} \frac{\rho(\br)\rho(\br')}{|\br - \br'|} .
\end{equation} 
$\cE_{xc}$ is the XC energy, approximated in practice as a semilocal functional of the density and its gradient, but can also include an explicit dependence on the orbitals (as in meta-generalized gradient functionals and hybrid functionals).

Excited states are obtained when the total energy is stationary with respect to the orbitals ${\bPsi} = \left(\ket{\psi_1}, \cdots, \ket{\psi_{M}}\right)^{T}$ 
with non-aufbau occupation numbers and can correspond to saddle points. The electronic energy surface has dimensionality 
$M N_b$, where $N_b$ is the number of grid points in an RSG or number of PW coefficients.
Even for a small molecule, the number of grid points can easily become large, on the order of $10^{6}$.
In order to facilitate the 
stationary-point search 
problem, the orbitals ${\bPsi} $ are expanded in a linear combination of some auxiliary orbitals $\bPhi$
\begin{equation}
{\bPsi} = U {\bPhi},
\end{equation}
where $U$ is an $M\times M$ unitary matrix. The auxiliary orbitals, $\bPhi$, can be 
chosen to be
the ground-state orbitals or 
any set of orbitals that represents an initial guess for the excited state. 
The energy can then be considered as a functional of both $U$ and $\bPhi$
\begin{equation}
\cE[\bPsi] = \cE[U\bPhi] = \cF[U, \bPhi] .
\end{equation}
Therefore, stationary points of $\cE[\bPsi]$ can be found in two steps: 
first by {\it extremizing} $\cF$ with respect to the expansion coefficients $U$ and then by {\it minimizing} the functional
\begin{equation}
\cL[\bPhi] = \underset{U} {\rm stat\,} \cF[U, \bPhi]
\end{equation}
with respect to $\bPhi$:
\begin{equation}
 \underset{\bPsi} {\rm stat\,} \cE[\bPsi]=\min_{\bPhi} \underset{U} {\rm stat\,} \cF[U, \bPhi]  = \min_{\bPhi} \cL[\bPhi] 
\end{equation}
The introduction of the additional functional $\cL[\bPhi]$
reduces the stationary-point search problem into two simpler tasks.
First, instead of finding a saddle point in a wave function space, one finds the saddle-point in the space of unitary matrices $U$ of low dimensionality. The reduction of dimensionality can further be achieved by decreasing the number of the virtual orbitals. For example, the first 
excited state of ammonia can be obtained by including only 4 virtual orbitals and, therefore, the dimensionality of the problem equals 24, within a frozen core approximation. 
This is a significant simplification as compared to the original problem of finding a saddle point in a $M N_b$ dimensional space.
Furthermore, an efficient algorithm based on a recently proposed~\cite{Levi2020} quasi-Newton method for finding saddle points in the space of unitary matrices can readily be used with minor modifications. 
The second simpler task is the outer loop minimization where conventional energy function minimization algorithms~\cite{NocedalWright} generalized for a wave function optimization can be employed.

The division of the original optimization problem into separate minimizations of different degrees of freedom
is a standard technique employed for ground-state calculations of metals~\cite{Gillan1989}, self-interaction corrected density functional calculations~\cite{Stengel2008,Klupfel2012,Lehtola2016a,Lehtola2018,Borghi2015a} and ensemble density functional calculations~\cite{Marzari1997a}. The inner and outer loops are further described below. The optimization of $\cE$ is further performed in conjunction with the maximum overlap method (MOM)~\cite{Gilbert2008} used to distribute the occupation numbers 
$\{f_m\}$, similar to what has previously been done in the context of LCAO~\cite{Levi2020fd,Levi2020}, and it is described in Appendix~\ref{appendix2}.

Additional considerations need to be addressed when a non-unitary invariant functional is used as in the case of 
PZ-SIC~\cite{Perdew1981}.
Perdew and Zunger proposed the following orbital-by-orbital correction~\cite{Perdew1981}
\begin{equation}
\cE^\mathrm{SIC}[\bPsi]  =\cE[\bPsi] - \sum_{i\sigma} \left( \cU[\rho_{i\sigma}] +  \cE_{xc}[\rho_{i\sigma}, 0]\right)
\end{equation}
making any approximation of the energy functional self-interaction free for one-electron systems.
The functional $\cE^\mathrm{SIC}$ is not invariant with respect to unitary transformations of the occupied orbitals and, therefore, 
an additional inner loop needs to be included in order to find the optimal orbitals in the occupied subspace 
minimizing 
the self-interaction corrected energy~\cite{Perdew1981,Pederson1984,Pederson1985a}.

In this case, the functional defined on the occupied subspace becomes unitary invariant~\cite{Stengel2008,Lehtola2014,Borghi2015a}. 
Thus, finding a solution that corresponds to the SIC excited state is achieved in a three-loop optimization:
\begin{equation}
    \underset{\bPsi} {\rm stat\,}  \cE^\mathrm{SIC} [\bPsi] = 
    \min_{\bPhi} \underset{U} {\rm stat\,}
    \underset{O} {\min}'\,
    \cE^\mathrm{SIC}[U O \bPhi] = \min_{\bPhi} \underset{U} {\rm stat\,} \cF^{SIC}[U, \bPhi]
\end{equation}
where the unitary minimization ${\min'\,}$ is performed among occupied orbitals only. 
More details on this additional inner loop are given in Appendix~\ref{appendix1}.


\subsection{Inner loop: Finding a stationary point of $\cF[U, \bPhi]$ with respect to $U$.}
 To find a stationary point of $\cF[U, \bPhi]$ for a given $\bPhi$, an exponential transformation is made,
 analogous to what has previously been done for energy minimization in wave-function based calculations~\cite{Rico1983a, Rico1983b, Douady1980, Head-Gordon1988} and density functional calculations~\cite{VanVoorhis2002,Lehtola2016a,Ivanov2020}, 
 and, more recently, for saddle-point searches ~\cite{Levi2020fd,Levi2020}. 
 The unitary matrix is parametrized as
\begin{equation}
U = e^{A}
\label{eq: innergrad}
\end{equation}
where $A$ is a skew-Hermitian matrix, $A^{\dagger} = -A$. The gradient of $\cF$ with respect to the elements of $A$ is
\begin{align}\label{eq: inner_gradient}
\frac{\partial \cF}{\partial A^{*}_{ij}} = \frac{2 - \delta_{ij}}{2} \left[ \int_0^1 e^{t A} L e^{-t A}  dt \right]_{ij},
\end{align}
where the elements of $L$ are given by
\begin{equation}
L_{ij} = \overline{\braket{\phi_j | \hat h_i | \phi_i}} - \braket{\phi_i | \hat h_j | \phi_j}
\end{equation}
 with $\hat h_j$ defined as
\begin{equation} \label{eq: gradE}
\hat h_j \ket{\psi_j} = \frac{\delta \cE}{\delta \bra{\psi_j}} = f_j  \left[ -\frac{1}{2} \nabla^2 + \hat v_{ext} + \hat v_{H} + \hat v_{xc} \right]\ket{\psi_j}
\end{equation}

 During the optimization, the elements of $A$ are found iteratively
 using a limited-memory version of the symmetric rank one quasi-Newton algorithm~\cite{Levi2020}. 
 The initial inverse Hessian is preconditioned with a diagonal matrix with elements~\cite{Head-Gordon1988}
\begin{align}\label{eq: prec1}
K_{ij} = \frac{1}{-2 (\epsilon_{i} - \epsilon_{j} ) (f_{i} - f_{j})}
\end{align}
where the $\epsilon_{i}$ are the eigenvalues of the KS Hamiltonian. 
Since this preconditioner is valid only for the canonical representation of the Hamiltonian, 
the auxiliary orbitals are updated to the canonical orbitals every $X$th iteration of the outer loop if the inner loop reaches a maximum number of iterations
\begin{equation}
 \bPhi \leftarrow C \bPsi = C U\bPhi,
\end{equation}
and set
\begin{equation}
 U = I
\end{equation}
where $C$ is the unitary matrix that transforms the auxiliary orbitals to the canonical orbitals. 
Further implementation details of the inner loop can be found in Refs.~\cite{Levi2020fd,Levi2020}


\subsection{Outer loop: Minimization of $\cL[\bPhi]$.}

Let $\mathcal{M}$  be a manifold in the Hilbert space such that
\begin{equation}
\mathcal{M} = \{\bPhi: \int d\br\, \phi_i^{*}(\br) \phi_j(\br)=\delta_{ij}, i, j=1\dots M \} .
\end{equation}
The tangent space to this manifold at $\bPhi$ is defined as 
 \begin{equation}\label{eq: tangentspace}
{\bV}_{\bPhi}({\bG}) = \lim_{\varepsilon\rightarrow0} \frac{\hat R[\bPhi + \varepsilon {\bG}] - \bPhi
}{\varepsilon} = \hat R_{\varepsilon}' [\bPhi + \varepsilon {\bG}]\big|_{\varepsilon=0}
\end{equation}
where $\hat R$ is the orthonormalization operator such that $ \hat R[\bPhi + \varepsilon {\bG}] \in \mathcal{M}$, 
and ${\bG}$ is a vector in the Hilbert space. 
For example, $\hat R$ can be chosen as the L\"owdin transformation.
Let $S_{\bX, \bY}$ be the overlap matrix between two vectors $\bX, \bY$ from the Hilbert space. Then
\begin{equation}\label{eq: overlap}
\begin{split}
 & S_{\bPhi + \varepsilon {\bm G}, \bPhi + \varepsilon {\bm G}} = \int d\br\, (\bPhi + \varepsilon {\bm G}) (\bPhi + \varepsilon {\bm G})^{\dagger} = \\ 
& I + \varepsilon \,\left(S_{\bPhi, {\bm G}} + S_{{\bG}, \bPhi}\right) + \varepsilon^{2} S_{{\bm G}, {\bm G}} = 
I + \varepsilon \,W_{\bPhi, {\bm G}} + \varepsilon^{2} S_{{\bm G}, {\bm G}}
\end{split}
\end{equation}
with $W_{\bPhi, {\bG}} = S_{\bPhi, {\bG}} + S_{{\bG}, \bPhi} $. 
For the L\"owdin transformation
\begin{equation}\label{eq: Roperator}
\hat R[\bPsi + \varepsilon {\bm G}] = S^{-1/2}_{\bPsi + \varepsilon {\bm G},\bPsi + \varepsilon {\bm G}} \, (\bPsi + \varepsilon {\bm G})
\end{equation}
and therefore, the tangent space at $\bPhi$ is
\begin{equation}\label{eq: tangentvector}
{\bV}_{\bPhi}({\bG}) = {\bG} - \frac{1}{2} W_{\bPhi, {\bG}}\, \bPhi ,
\end{equation} 
obtained after substituting eqs~\eqref{eq: overlap}~and~\eqref{eq: Roperator} into 
eq~\eqref{eq: tangentspace} keeping only first order terms with respect to $\varepsilon$.

The gradient of $\cL$ can be calculated as:
\begin{equation}\label{eq: gradL}
\frac{\delta \cL}{\delta \bPhi^{*}} =U \frac{\delta \cE}{\delta \bPsi^{*}}, 
\end{equation}
where 
\begin{equation} \label{eq: gradE2}
\left(\frac{\delta \cE}{\delta \bPsi^{*}}\right)_{j} = \frac{\delta \cE}{\delta \bra{\psi_j}}
\end{equation}

After defining the tangent space in equation~\eqref{eq: tangentspace} and the gradient in equations~\eqref{eq: gradL} and \eqref{eq: gradE2}, 
the minimization of $\cL$ can be written as

\vskip 0.3 true cm
{\bf Minimization algorithm.}
\begin{itemize}
\item Set $k \gets 0$, choose initial guess ${\bPhi}^{(k)}$; calculate gradient $\bG^{(k)} = \delta \cL/\delta \bPhi^{*(k)}$ using eqs~\eqref{eq: gradL} and \eqref{eq: gradE2}.
\item Project gradient on the tangent space ${\bV^{(k)}} = {\bV}({\bG^{(k)}})$ at ${\bPhi}^{(k)}$.
$\varepsilon$ and calculate residual error $\Delta^{(k)} = \| {\bV^{(k)}} \|$.
\item While $\Delta^{(k)} > \varepsilon$:
\begin{enumerate}
\item Compute search direction $\bP^{(k)}$ according to the chosen minimization algorithm and apply preconditioning (for example, for gradient descent, $ \bP^{(k)} = - \bG^{(k)}$ and inverse kinetic energy operator as preconditioner~\cite{Briggs1995,Kresse1996prb}).
\item Project the search direction on the tangent space ${\bV^{(k)}} = {\bV}(\bP^{(k)})$ at ${\bPhi}^{(k)}$.
\item  Choose optimal step length $\alpha^{(k)}$ along $\bV^{(k)}$ and compute 
\begin{equation}\label{eq: iteratewfs}
{\bPhi}^{(k+1)} \gets {\bPhi}^{(k)} + \alpha^{(k)} {\bV^{(k)}}
\end{equation}
\item Orthonormalize the wave functions, $ {\bPhi}^{(k+1)} \gets \hat R[{\bPhi}^{(k+1)}]$.
 
\item Compute new gradient $\bG^{(k+1)}$ and project it on \\ the tangent space ${\bV^{(k+1)}} = {\bV}({\bG^{(k+1)}})$ at ${\bPhi}^{(k+1)}$.

\item  Calculate residual $\Delta^{(k+1)} = \| {\bV^{(k+1)}} \|$.
 \item $k \leftarrow k+1$.
\end{enumerate}
\item End.
\end{itemize}

The search direction $\bP^{(k)}$ can be chosen using a conjugate gradient~\cite[p.~121]{NocedalWright} or a limited memory quasi-Newton  algorithm. Here, the limited-memory Broyden-Fletcher-Goldfarb-Shanno (L-BFGS) algorithm as described in Ref.~\cite[p.~177]{NocedalWright} is used. For minimization, this algorithm is known to give fast and robust convergence. 

\section{Implementation and Numerical Tests}\label{sec: numtest}

The DO-MOM algorithm has been implemented in a development branch of 
the Grid-based projector augmented wave (GPAW)
software~\cite{GPAW2} and can be used with either a finite-difference RSG~\cite{GPAW1} or PW basis set. The calculations are carried out using the frozen core approximation and the projector-augmented wave method~\cite{PAW_Blochl:1994}. The iterative SCF algorithms used here 
are based on either the Davidson algorithm~\cite{Davidson1975} or the RMM-DIIS algorithm~\cite{Kresse1996prb} as implemented in GPAW. 
Both versions of the SCF algorithm make use of Pulay density mixing~\cite{Pulay1980} and MOM~\cite{Gilbert2008} (see~Appendix~\ref{appendix2}).  
Default values of the convergence parameters are used. The Pulay density mixing uses densities from three previous steps and the 
coefficient used in the linear mixing of the density with the density residual vector is 0.15. No damping of short-wavelength density changes is used~\cite{GPAW2}. 
In the DO-MOM calculations, Pulay density mixing is not used. Instead, the density is calculated from the orbitals obtained at each
iteration. 
The gradient of the energy projected on the tangent space in the outer loop in DO, or the residual vector in SCF, is preconditioned with the inverse kinetic energy operator~\cite{Briggs1995,Kresse1996prb}.

In the outer loop of the DO-MOM algorithm, the search direction is calculated according to the L-BFGS algorithm, 
using only the previous step to estimate the Hessian matrix. For a quasi-Newton algorithm a step length of 1 is a natural choice. 
However, the first step in the optimization corresponds to the gradient descent algorithm and the following maximum step length update is used
in order to avoid too large changes in the orbitals: 
if the norm of the search direction
\begin{equation}
\Delta = \bP^{\dagger} \bP
\end{equation}
is larger than $\alpha_{max}$ ($\Delta > \alpha_{max}$) then
\begin{equation}
\bP \gets \frac{\alpha_{max}}{\Delta}\bP .
\end{equation}
The value $\alpha_{max}=0.25$ is found to give reliable convergence.
For the inner loop optimization, the limited-memory symmetric rank one update~\cite{Levi2020} is used.

The calculations were carried out in the following way if not stated otherwise: The molecule is placed in a rectangular box with at least 7 {\AA} vacuum space in 
all directions from the nuclei to the boundary of the box. 
Open boundary conditions are used. 
A grid mesh spacing of 0.2 {\AA} is employed. All the calculations are spin-unrestricted and use the PBE functional~\cite{Perdew1996}.

\begin{figure}[H]
\centering
\includegraphics[width=0.6\textwidth]{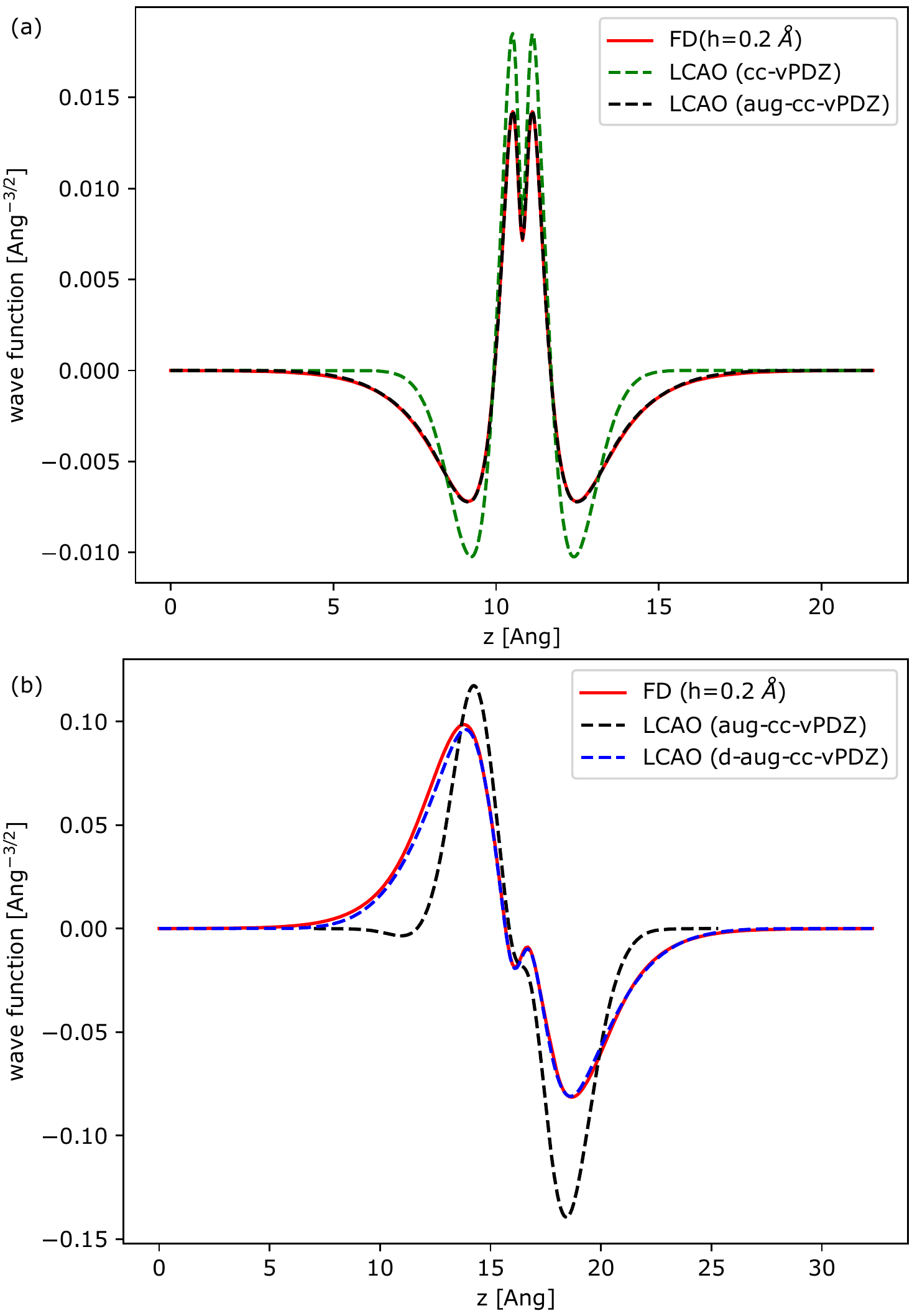}
\caption{
Comparison of three basis sets in a PBE calculation of the (a) 3$s$ and (b) 3$p_z$ Rydberg states of NH$_3$ for excitation from the HOMO. Both orbitals 
are shown along 
the rotational axis of the molecule ($z$-axis). 
The finite-difference real-space grid calculation with grid spacing of 0.2 {\AA} (red)
is closely reproduced by an LCAO calculation when the 
aug-cc-pVDZ basis set (black dashed) is used for 3s excitation while employing  cc-pVDZ (green dashed) is not enough. However,
an expanded d-aug-cc-pVDZ basis set including an extra set of diffuse functions is required (blue dashed) for 3$p_z$, as  
the aug-cc-pVDZ basis set is clearly not sufficient.
} 
\label{fig:Ammonia3pz}
\end{figure}

The advantage of RSG over LCAO is illustrated with a
calculation of the 3$s$ and 3$p_z$ Rydberg states of NH$_3$ for excitation from the HOMO, see
Fig.~\ref{fig:Ammonia3pz}. For the 3$s$ excitation, an LCAO calculation using a cc-pVDZ basis set is clearly not sufficient to reproduce the RSG results. Using a aug-cc-pVDZ basis set, which includes diffuse functions, results in a closer agreement between RSG and LCAO. For an excitation to the more diffuse 3$p_z$ Rydberg orbital, an expanded d-aug-cc-pVDZ basis set, including additional diffuse functions, is needed. The RSG approach is more flexible for such calculations and does not require changes in the basis set approximation.

The energy of an open-shell singlet state is calculated using
the spin purification~\cite{Ziegler1977}:
\begin{equation}\label{eq: SpinPur}
    E^{s} = 2E(\uparrow\downarrow) - E(\uparrow\uparrow),
\end{equation}
where $E(\uparrow\uparrow)$ is the energy of the triplet state and $E(\uparrow\downarrow)$ is the energy of the mixed spin state. Both states are calculated independently and variationally.
The singlet excited-state energy will hereafter be referred to as the energy calculated according to eq~\eqref{eq: SpinPur}.
This has been found to give a better estimate 
of the singlet excited-state energy
in eDFT calculations using semilocal KS functionals compared to the estimate obtained from the mixed spin state~\cite{Kowalczyk2011}. 

An excited state calculation is initialized by swapping occupation numbers corresponding to the targeted excitation (i.e. from HOMO to LUMO+1), between the 
orbitals obtained from the ground-state calculation. 
An exception to this is in the study of the donor-to-donor charge transfer electronic excitation 
in the water dimer, which is initialized from the orbitals obtained from a separate 
excited-state calculation describing a donor-to-acceptor charge-transfer excitation.

\subsection{Test I: Charge-transfer excitations in nitrobenzene}

Charge-transfer excitations in nitrobenzene are known to be challenging cases for conventional algorithms~\cite{Mewes2014, Hait2020} 
and are often used as benchmark tests~\cite{Hait2020, Levi2020fd, Levi2020, CarterFenk2020}. 
The {$^1$}A$_1$($\pi^\prime \rightarrow \pi^*$) excitation transfers an electron from the benzene ring to the nitro group while in
the {$^1$}A$_1$(n$_\pi \rightarrow \pi^{\prime*})$ excitation the transfer is in the opposite direction. 

A ground-state calculation 
including 9 virtual orbitals
was first performed to obtain the initial orbitals.
The {$^1$}A$_1$($\pi^\prime \rightarrow \pi^*$) excited state 
was then calculated by promoting an electron from the HOMO-2 to the LUMO 
orbital, while the {$^1$}A$_1$(n$_\pi \rightarrow \pi^{\prime*}$) state 
was calculated by promoting an electron from the HOMO-4 to the LUMO+1 
orbital. 

\begin{figure}[H]
    \centering
    \includegraphics[width=1.0\textwidth]{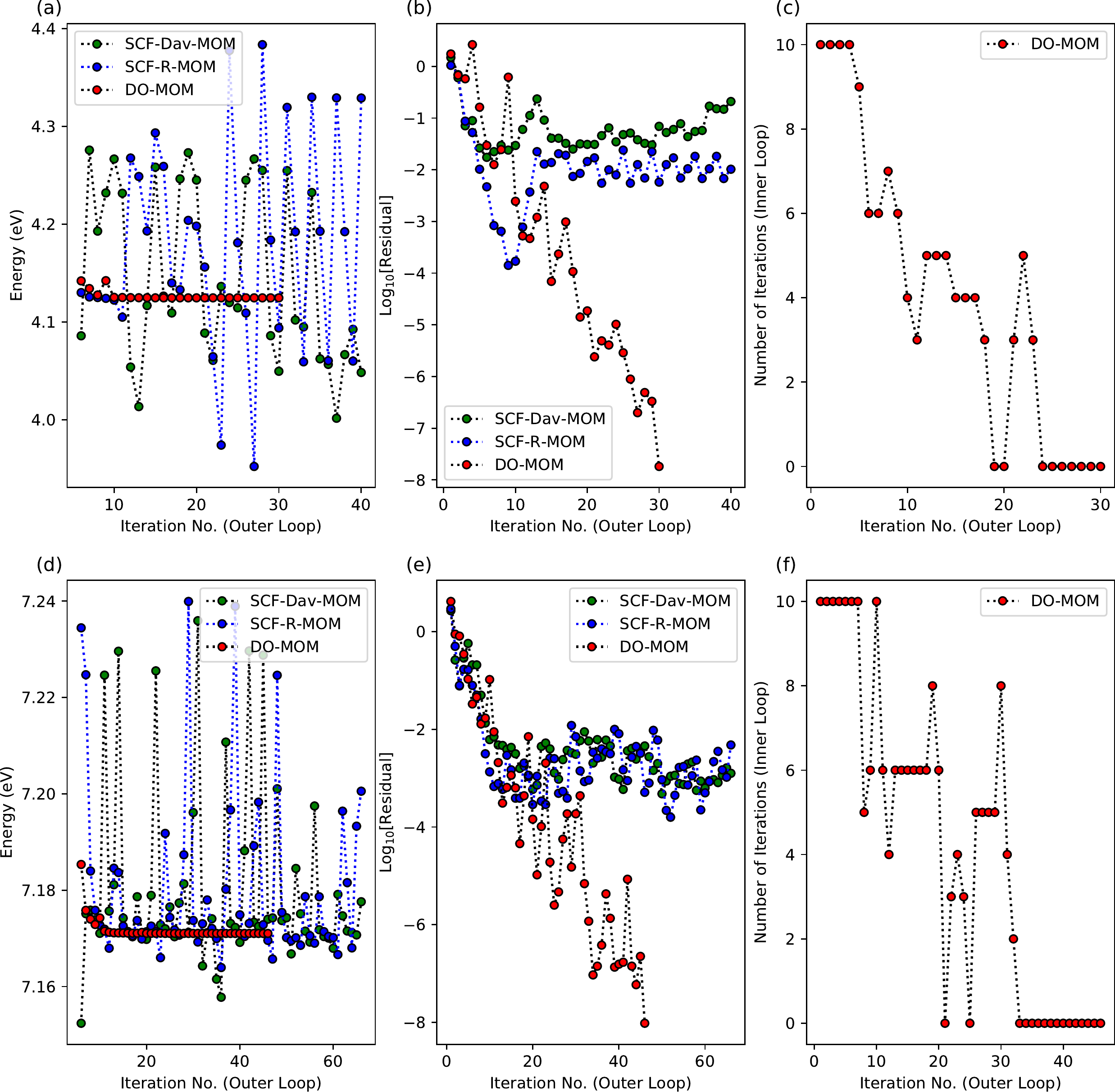}
    \caption{
    Comparison of the performance of various algorithms in calculations of the {$^1$}A$_1$($\pi^\prime \rightarrow \pi^*$) excitation
    (a,b,c) and {$^1$}A$_1$(n$_\pi \rightarrow \pi^{\prime*})$ excitation (d,e,f) of nitrobenzene.
    The SCF-MOM method, based either on Davidson (SCF-Dav-MOM) or RMM-DIIS (SFC-R-MOM) algorithms, does not converge well, while DO-MOM gives robust convergence. 
    (a) and (d) show the convergence of the energy; (b) and (e) show the norm of the residual, and  
    (c) and (f) show the number of inner-loop iterations for each outer-loop iteration in the DO-MOM calculation.
     } 
    \label{fig: Nitrobenzene}
\end{figure}

 An analysis of the performance of the DO-MOM calculation and comparison with the two versions of the SCF-based methods
 is presented in Fig.~\ref{fig: Nitrobenzene}. Both the Davidson and RMM-DIIS implementations of SCF-MOM quickly approach the excitation energy of the target solution but then show erratic behaviour. 
 This is attributed to the presence of several orbitals with energy close to that of the orbital from which excitation occurs~\cite{Levi2020fd,Levi2020}. 
The energy difference between HOMO-4 and HOMO-1 is only 0.56 eV and a change in the
 ordering of the orbitals occurs during the optimization of the excited state~\cite{Levi2020fd,Levi2020}. 
 It is known that for orbitals that are energetically close, SCF algorithms have a difficulty converging unless smearing of 
 occupation numbers is used
 or the parameters in the Pulay mixing are fine tuned. In contrast, the DO-MOM algorithm shows robust convergence. 
 Tight convergence is obtained within 30 to 45 outer loop iterations as shown in Fig.~\ref{fig: Nitrobenzene}(b)~and~(e). 
 Initially, during each outer loop 
 iteration, several inner loop 
 iterations are performed as shown in Fig.~\ref{fig: Nitrobenzene}(c)~and~(f). 
 Towards the minimum of the energy functional $\cL$,
 no inner-loop iterations 
 are performed,
 only outer-loop iterations.

\subsection{Test II: Excitation to degenerate orbitals}

Degenerate electronic states 
need to be
represented by multi-determinant wave functions. 
When a single determinant is used in a KS-DFT calculation two problems occur. 
The first is a technical problem, as the SCF algorithm 
has difficulty converging 
unless large enough smearing of the occupation numbers is used. 
The second is a conceptual problem, in that different single determinants which should in principle be degenerate can 
give different electron densities and, as a result, different total energy. 
This 
occurs, for example, 
in calculations of open-shell atoms~\cite{Becke2002, Johnson2007}. 
In the case of degenerate excited states, an additional problem arises. In order to unambiguously assign an excited state, 
the 
wave function needs to have the symmetry of the excited state.
With real-valued orbitals, which are most commonly employed in electronic structure calculations, 
this requirement is not necessarily satisfied due to symmetry breaking, as 
is demonstrated below.

\begin{figure}[H]
    \centering
    \includegraphics[width=1.0\textwidth]{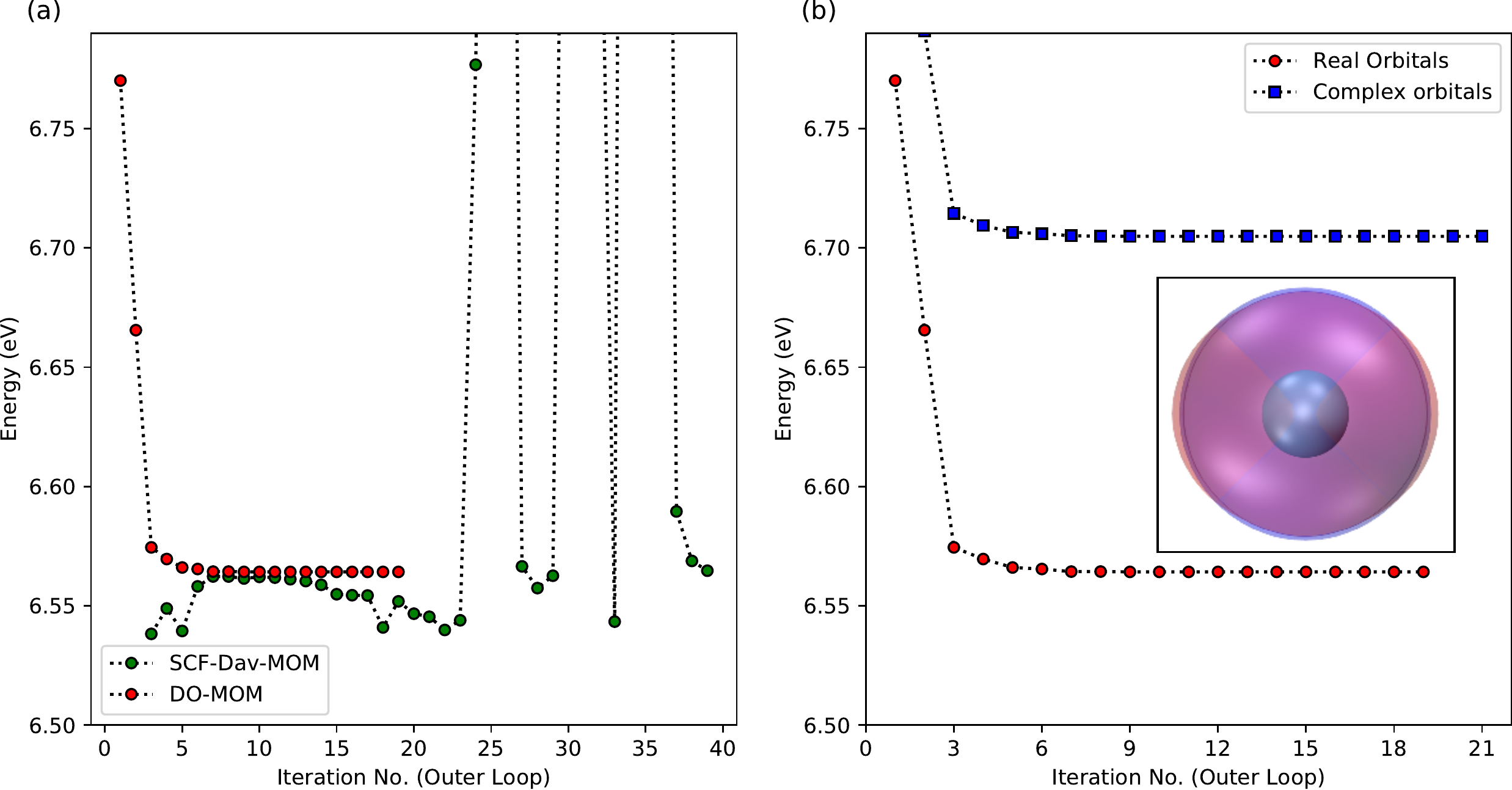}
    \caption{(a) Comparison of the performance of the SCF-MOM method implemented with Davidson and the 
    DO-MOM method in calculations of the lowest n$\rightarrow\pi^{*}$ excitation of a carbon monoxide molecule. 
    (b) Comparison of the DO-MOM calculations with real and complex orbitals. 
    The inset shows the total electron density obtained with real orbitals (red) and complex orbitals (blue). 
    The internuclear axis is perpendicular to the plane of the image. The density obtained using real orbitals 
    lacks the uniaxial symmetry.} 
    \label{fig: CO}
\end{figure}

Consider the lowest valence excited state in carbon monoxide. Using a single determinant, this excited state can be described by the promotion of an electron from a $\sigma$ orbital (ground-state HOMO) to one of the two lowest degenerate $\pi^{*}$ orbitals, 
($\pi^{*}_{x}$ or $\pi^{*}_{y}$, the ground-state LUMO, in the case of real wave functions). 
For such an excitation, the SCF method with the Davidson algorithm 
does not converge with integer occupation numbers. 
The energy oscillates around the excited-state solution, as shown in Fig.~\ref{fig: CO}(a). 
The DO-MOM algorithm, however, gives smooth convergence to the excited-state solution.

Since the orbitals are real, they are not eigenstates of the $z$-component of the angular momentum operator, 
and the angular momentum of the single-determinant wave function around the internuclear axis ($z$-axis) is not defined. 
In addition, the resulting electron density lacks uniaxial symmetry. It has instead an elliptic shape
in the 
$x$-$y$ plane 
with orientation
depending on which orbital is occupied, $\pi^{*}_x$ or $\pi^{*}_y$, see Fig.~\ref{fig: CO}(b).
This is inconsistent with the symmetry of the molecule. 
In the DO-MOM calculation, the orbitals can be chosen to be complex valued functions without any modifications of the algorithm.
If the LUMO is chosen as a complex $\pi^{*}_{+1}$ or $\pi^{*}_{-1}$ orbital, where +1 or -1 is the eigenvalue of the $z$-component angular momentum operator, the single-determinant excited-state wave function has a well-defined angular momentum and can unambiguously 
be identified as a $\Pi$ state.
The solution DO-MOM converges to using complex orbitals has 0.15 eV higher energy compared to the real-valued solution, 
but it is more accurate since the 
total density then has rotational symmetry around the internuclear axis [see Fig.~\ref{fig: CO}(b)]. 
Thus, the use of complex orbitals not only allows one to properly represent the total angular momentum of the excited state, 
but it also provides a density with the correct symmetry. The spin symmetry is still broken, however, in the unrestricted calculation. 

The importance of using complex orbitals in order to provide correct description of the ground state 
has been emphasized in calculations of atoms and molecules using self-interaction corrected functionals~\cite{Klupfel2011,Klupfel2012mol,Lehtola2016b} 
as well as 
within restricted
Hartree-Fock theory~\cite{Small2015} and 
KS formalism~\cite{Lee2019}. In particular, in the work of Lee {\it et. al}~\cite{Lee2019} it was shown that real orbitals break the 
cylyndrical symmetry of the density in the singlet ground state of 
O$_2$ while complex orbitals restore such symmetry within the restricted KS formalism. 
Here, it is shown that a similar situation occurs in the excited states of open-shell singlets 
and that the symmetry can be restored in the spin-unrestricted formalism using complex orbitals.

\section{Application I: Excited states of an Fe$^{\rm II}$ carbene photosensitizer}\label{sec: application}

\begin{figure}[H]
\centering
\includegraphics[width=1.0\textwidth]{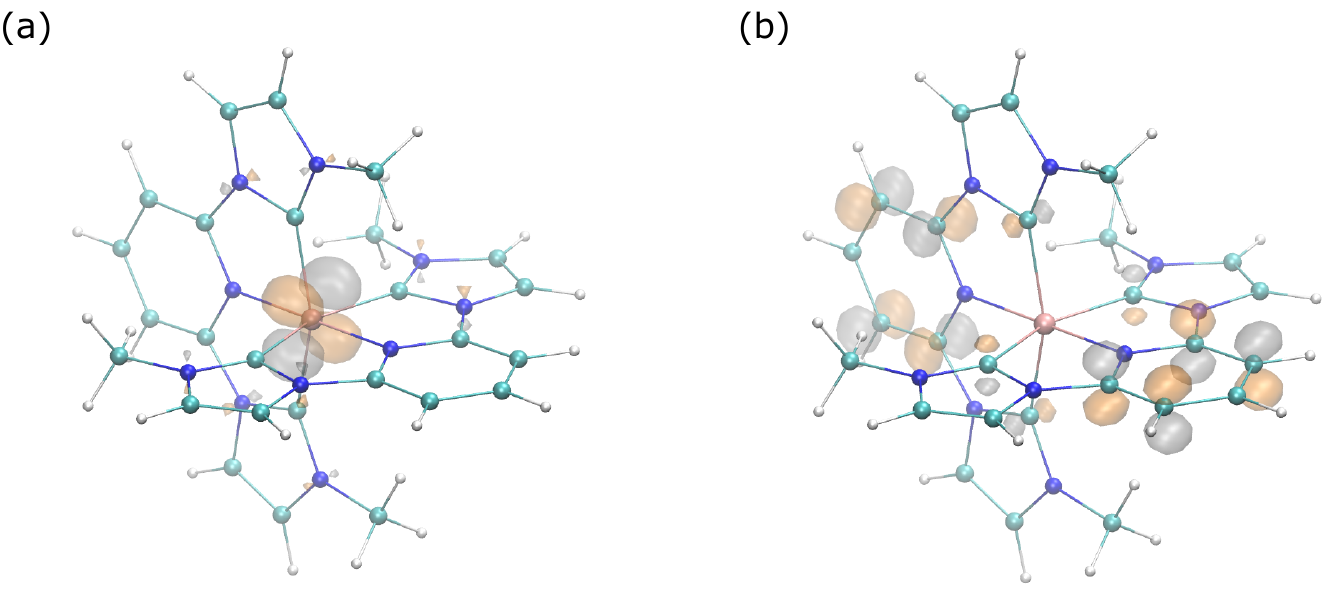}
\caption{
Atomic structure of the [Fe(bmip)$_2$]$^{2+}$ complex together with the orbitals  involved in the transition to the bright MLCT state
(C atoms green, N atoms blue, H atoms white, Fe atom orange).  
(a) HOMO orbital of the ground state localized on the Fe atom. 
(b) LUMO+2 orbital of the ground state delocalized over the ligands. Promotion of the electron from HOMO to LUMO+2 corresponds to a charge-transfer excitation (denoted as ${}^1$MLCT$_2$ state in the text). 
The isosurfaces are shown with grey and orange colors and 
correspond to isovalues of $\pm0.16$ {\AA}$^{-3/2}$.
} 
\label{fig: FecomplexOrbitals}
\end{figure}

The first application of the DO-MOM method involves calculations of four MLCT and two MC excited states of the 
[Fe(bmip)$_2$]$^{2+}$ complex that consists of 63 atoms (see Fig. \ref{fig: FecomplexOrbitals}).
The calculations are carried out with the BLYP functional~\cite{Becke1988,LYP1988}. The ground-state geometry was chosen to be the same as the geometry optimized with the B3LYP$^{*}$ functional~\cite{Reiher2001} in Ref.~\cite{Papai2016}. In the DO-MOM calculations, point-group symmetry constraints for the total electron density are used.

\begin{figure}[H]
    \centering
    \includegraphics[width=1\textwidth]{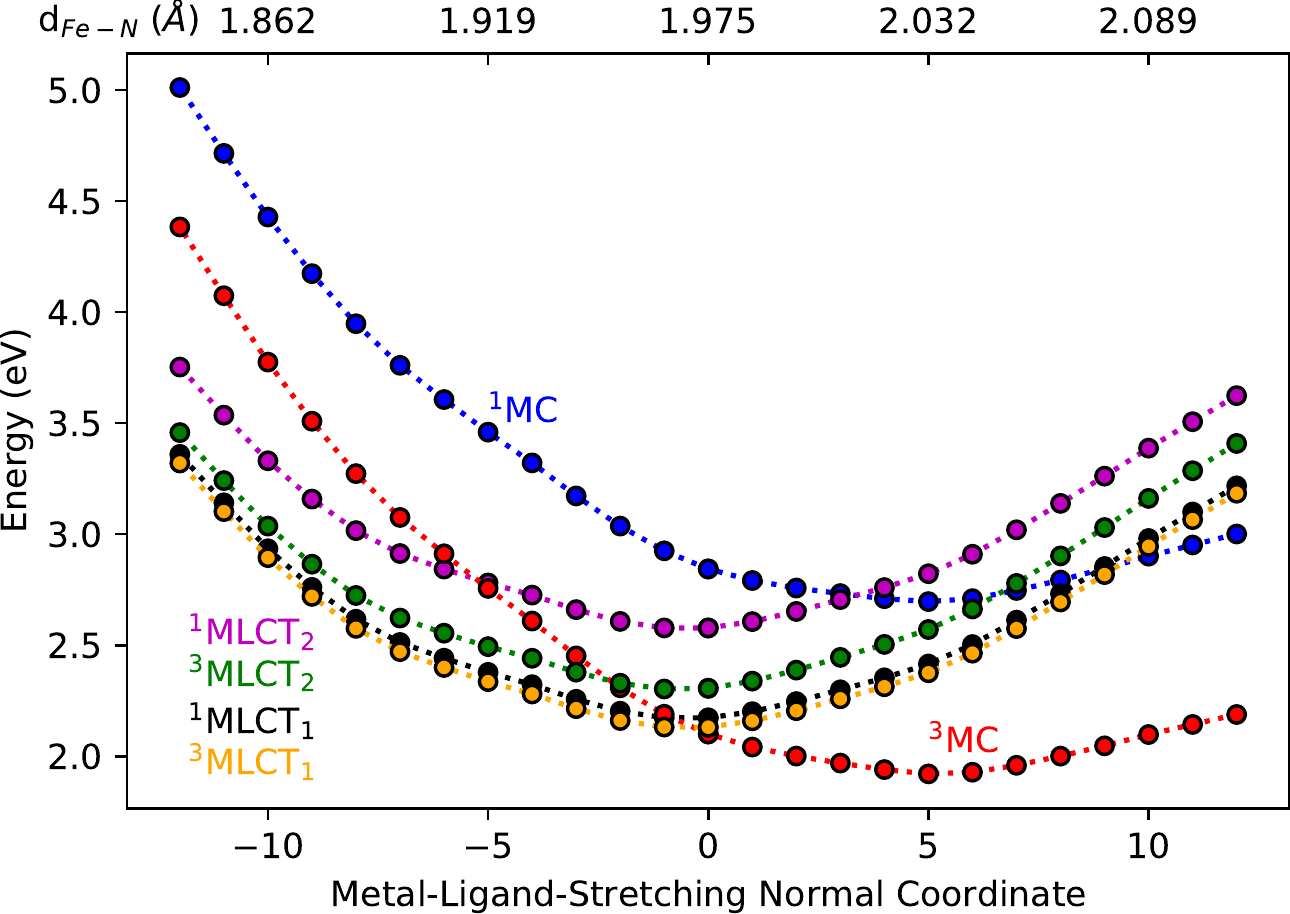}
    \caption{Potential energy curves along a metal-ligand normal coordinate (the breathing mode $Q_6$ as defined in Ref.~\cite{Papai2019a}) that is expected to be responsible for the nuclear dynamics during the relaxation from the initially photoexcited $^1$MLCT state to the ${}^{3}$MC state. MLCT: metal-to-ligand charge-transfer excitation, MC: metal-centered excitation. The zero of the lower x-axis corresponds to the ground-state B3LYP$^*$ optimized geometry~\cite{Papai2016}. The upper x-axis shows the shortest iron-nitrogen pair distance in Å.
    } 
    \label{fig:FeCarbene}
\end{figure}

%
Fig.~\ref{fig:FeCarbene} shows the
energy of the various states calculated along the metal-ligand bond stretching coordinate (the $Q_6$ breathing normal mode according to Ref.\cite{Papai2019a}) that is 
believed
to account for the nuclear dynamics following photoexcitation~\cite{Papai2016,Kunnus2020}. 
The singlet state labelled $^1$MLCT$_2$ in Fig.~\ref{fig:FeCarbene}, corresponding to a HOMO-to-LUMO+2 transition (see Fig.~\ref{fig: FecomplexOrbitals}), has a vertical excitation energy of 2.58 eV, only 0.13 eV lower than the position of the maximum of the experimental UV/VIS absorption spectrum of the complex dissolved
in acetonitrile~\cite{Liu2013}. Indeed, this state has the same character as the 
state with largest oscillator strength in TDDFT calculations~\cite{Papai2019a};
thus, confirming that $^1$MLCT$_2$ corresponds to the bright MLCT state with a calculated excitation energy in good agreement with experiment.
The triplet with same orbital occupancy (labelled $^3$MLCT$_2$ in Fig.~\ref{fig:FeCarbene}), 
the lower-lying singlet ($^1$MLCT$_1$) and triplet MLCT ($^3$MLCT$_1$) states arising from the HOMO-to-LUMO excitation 
are also shown, as well as
the lowest triplet MC state and the corresponding singlet with same character (arising from HOMO-1 to LUMO+4 
excitation).
The orbitals involved in the transitions to these states are shown in the Supporting Information.


The combined ultrafast X-ray emission and scattering experiments have
detected vibrational wavepacket dynamics along the metal-ligand stretching coordinate in a $^3$MC state with a period of 278 fs\cite{Kunnus2020}. 
The curvature of the energy curve for the $^3$MC state calculated with DO-MOM is used to estimate the 
vibrational period, obtaining a value of 280 fs (see Supporting Information).
The shape of the PES predicted by DO-MOM with BLYP, therefore, appears to agree well with the 
experiment in this respect, lending support for the use of DO-MOM in future dynamics simulations to study
this and other photosensitizer complexes.

\section{Application II: Assessment of Perdew-Zunger \\ self-interaction correction}\label{sec: PZ-SIC}

Calculations are carried out for 9 valence and 9 Rydberg excitations in 13 molecules involving both singlet and triplet 
excited states. The results of the calculations are compared to theoretical best estimates from Ref.\cite{Loos2018}, 
which include corrections for basis set limitations and "all-electron" effects.
The "all-electron" relaxation effects are estimated to be small, around $0.01-0.02$ eV~\cite{Loos2018}, in the present cases. 
On the other hand, the basis set correction can be significant, 
and is important for a consistent comparison with the finite-difference RSG results obtained in the present work. 
The molecules are placed in a rectangular box with at least 9 {\AA} vacuum space in 
in all directions from 
any of the nuclei 
to the edge of the simulation box.
This is found to be large enough to correctly describe even the diffuse Rydberg orbitals. 
The atomic coordinates of the molecules are 
those given in Ref.~\cite{Loos2018}. The number of virtual orbitals is set to 8. 
The excitation energy is calculated with respect to the energy of the singlet ground state. 
No symmetry constraints are enforced on the total density.
The SIC calculations are carried out using the PBE functional with full PZ-SIC (PBE-SIC) and with the PBE functional where 
the correction is scaled by a half (PBE-SIC/2) as such scaling has previously been found to provide better estimates of 
atomization energy of molecules and band gaps of solids~\cite{Klupfel2012mol,Jonsson2011a}.

The DO-MOM calculated values of the excitation energy 
are given in Table~\ref{tbl: triplets} for the triplet states and in Table~\ref{tbl: pure-singlets} for the singlet states.
Since mixed spin states are often used in practice as an approximation to a singlet energy surface, 
such calculations are also performed and the results are given in Table~S1 in the Supporting Information.

The calculations using the PBE functional give a mean error (ME) of -0.27 eV 
and a root mean square error (RMSE) of 0.31 eV with respect to the theoretical best estimates for the 
excitations to triplet states while a larger error is obtained for excitations to singlet states, ME of -0.46 eV and RMSE of 0.54 eV. 
If the spin purification for singlet states is not applied, the error in the excitation energy 
is significantly larger, with the RMSE being 0.95 eV.

The calculations using self-interaction correction, i.e. the PBE-SIC functional,
give slightly more accurate values of the excitation energy for the singlet excited states, the magnitude of the ME with respect to the theoretical best estimates  
being 0.2 eV smaller 
compared to PBE while the MAE is not improved as much. For the triplet excited states, PBE-SIC does not lead to an improvement in MAE as compared to PBE but this is largely due to a few outliers in the data set such as diazomethane where the excitation energy with PBE-SIC is underestimated by around 0.9 eV as compared to 0.5 eV obtained with PBE. Generally, we observe that the excitation energy is improved when calculated with PBE-SIC as compared to PBE when the self-interaction error is dominant in the excited state, compared to the ground state.

As has been shown previously in ground-state calculations, it is important to use complex orbitals in SIC calculations~\cite{Klupfel2011,Klupfel2012mol,Lehtola2016b}.
This is also found to be the case here in the calculated values of the excitation energy. 
If real orbitals are used, the calculated values of the excitation energy become worse than those obtained with the PBE functional
(the RMSE being 0.44 eV for triplet excitations, see Table~S4 in Supporting Information).

For the triplet excitations, the scaled self-interaction correction, PBE-SIC/2 functional, gives smaller improvement
while for the singlet excitations the MAE and RMSE with respect to the theoretical best estimates is a bit smaller than for full correction PBE-SIC. For the mixed spin states, PBE-SIC performs better (see Table S1 in Supporting Information).

While the mean errors for the whole data set are fairly similar between PBE and PBE-SIC calculations, the performance of these functionals for valence excitations is different than for Rydberg excitations. For valence triplet excited states, PBE-SIC, scaled or not, performs worse than PBE on average while the opposite trend is observed for Rydberg excitations~(see Figs.~S5~and~S6 in the Supporting Information). Furthermore, PBE-SIC does not affect the excitation energy in the same way as exact exchange. Calculations with PBE0 (25\% of exact exchange) and PBE50 (50\% of exact exchange) systematically reduce the excitation energy of valence triplet states as compared to the PBE functional~(see Fig.~S5~and~S6 and Table~S2 in Supporting Information), while PBE-SIC/2 and PBE-SIC increase the excitation energy in cases where the SIC in the excited state is larger than that in the ground state. PBE-SIC/2 performs better for triplet excitations than PBE0 and PBE50, and shows similar performance for the singlet excited states (see Tables~S2~and~S3 in Supporting Information).

PZ-SIC sometimes has a small effect on the excitation energy even in cases where the spatial extent of the excited electron and the hole are very dissimilar. Orbital-by-orbital estimate of the self-interaction energy, which is most appropriate for a single electron system,
turns out to be of similar magnitude for the ground and excited states.
This is a surprising result since the classical self-Coulomb energy of a diffuse, Rydberg orbital is known to be smaller than that of a  
more localized ground-state orbital. Table~\ref{tbl: si_water} shows an analysis of this for the water molecule.
A near cancellation of the total self-interaction energy still occurs because there is a simultaneous change in the self-XC term 
that offsets the difference in the classical self-Coulomb energy.

\begin{table}[H]
\scriptsize
  \caption{
Energy of excitations to triplet states calculated with the DO-MOM method and comparison with theoretical
best estimates as well as experimental values. The calculations make use of a generalized gradient
approximation Kohn-Sham functional (PBE), with scaled self-interaction correction (SIC/2) and
with full self-interaction correction (SIC). The mean error (ME), mean absolute error (MAE) and
root mean square error (RMSE) are given with respect to theoretical best estimates and with 
respect to experimental values at the bottom of the table.
  }
  \label{tbl: triplets}
  \begin{tabular}{ l  c  c  c c c c }
    \hline
    molecule & excitation &  PBE & SIC/2 & SIC &  TBE\textsuperscript{\emph{a}} & EXP\textsuperscript{\emph{b}} \\
    \hline
acetaldehyde & 1$^3$A$^{\prime\prime}$ ($n\rightarrow\pi^*$; V) & 3.65 & 3.75 & 3.79 & 3.98 & 3.97 \\
\\
acetylene & $1^3\Delta_u$($\pi\rightarrow\pi^*$; V) & 6.33 & 5.89 & 6.04 & 6.40 & 6.0 \\
\\
ammonia & 2$^3$A$_1$(n$\rightarrow$3s; R) & 6.16 & 6.10 & 6.06 & 6.37 & 6.02 \\
\\
carbon monoxide & $1^3\Pi$(n$\rightarrow\pi^*$; V) & 5.91 & 5.84 & 5.66 & 6.28 & 6.32 \\
\\
diazomethane & 1$^3$A$_2$($\pi\rightarrow\pi^*$; V) & 2.76 & 2.38 & 1.88 & 2.80 &         \\
\\
\\
ethylene & 1$^3$B$_{3u}$(n$\rightarrow$3s; R) & 7.01 & 7.05 & 7.07 & 7.28 & 6.98 \\
  & 1$^3$B$_{1u}$($\pi\rightarrow\pi^*$; V) & 4.46 & 4.63 & 4.75 & 4.54 & 4.36 \\
\\
formaldehyde & 1$^3$B$_2$(n$\rightarrow$3s; R) & 6.69 & 6.99 & 7.12 & 7.14 & 6.83 \\
\\
formamide & 1$^3$A"(n$\rightarrow\pi^*$; V) & 5.14 & 5.23 & 5.27 & 5.37 & 5.2 \\
\\
hydrogen sulfide & 1$^3$A$_2$(n$\rightarrow$4p; R) & 5.39 & 5.44 & 5.43 & 5.74 & 5.8 \\
\\
ketene & 1$^3$B$_1$($\pi\rightarrow$3s; R) & 5.64 & 5.77 & 5.79 & 5.85 & 5.8 \\
\\
methanimine & 1$^3$A"(n$\rightarrow\pi^*$; V) & 4.20 & 4.35 & 4.41 & 4.64 &         \\
\\
thioformaldehyde & 1$^3$A$_2$(n$\rightarrow\pi^*$; V) & 1.71 & 1.81 & 1.88 & 1.94 &         \\
  & 1$^3$B$_2$(n$\rightarrow$4s; R) & 5.31 & 5.54 &  5.67 & 5.76 &         \\
  & 2$^3$A$_1$($\pi\rightarrow\pi^*$; V) & 3.36 & 3.33 & 3.28 & 3.44 & 3.28 \\
\\
water molecule& 1$^3$B$_1$(n$\rightarrow$3s; R) & 7.10 & 7.09 & 7.08 & 7.33 & 7.2 \\
  & 1$^3$A$_2$(n$\rightarrow$3p; R) & 8.75 & 8.87 & 8.97 & 9.30 & 8.9 \\
  & 2$^3$A$_1$(n$\rightarrow$3s; R) & 9.28 & 9.25 & 9.23 & 9.59 & 9.46 \\
\\
\hline
\rowcolor{Gray}
& ME (TBE) & -0.27 & -0.25& -0.25 & & \\
& ME (EXP) & -0.09 & -0.06 & -0.04 & &  \\
\rowcolor{Gray}
& MAE (TBE) & 0.27 & 0.26 & 0.27 & &  \\
& MAE (EXP) & 0.19 & 0.16 & 0.18 & &  \\
\rowcolor{Gray}
& RMSE (TBE) & 0.31 & 0.29 & 0.34 & & \\
& RMSE (EXP) & 0.22 & 0.21 & 0.26 & & \\
\end{tabular}

  \textsuperscript{\emph{a}}Theoretical best estimates as given in Ref.~\cite{Loos2018}. \ 
  \textsuperscript{\emph{b}}Experimental values listed in Ref.~\cite{Loos2018} (see references therein).
\end{table}


\begin{table}[H]
\scriptsize
  \caption{
Energy of excitations to singlet states (spin purified) calculated with the DO-MOM method and comparison with theoretical
best estimates as well as experimental values. The calculations make use of a generalized gradient
approximation Kohn-Sham functional (PBE), with scaled self-interaction correction (SIC/2) and
with full self-interaction correction (SIC). The mean error (ME), mean absolute error (MAE) and
root mean square error (RMSE) are given with respect to theoretical best estimates and with 
respect to experimental values at the bottom of the table.
  }
  \label{tbl: pure-singlets}
  \begin{tabular}{ l  c  c  c  c c c}
    \hline
    molecule & excitation & PBE & SIC/2 &
    SIC &
    TBE\textsuperscript{\emph{a}} & EXP\textsuperscript{\emph{b}} \\
    \hline
acetaldehyde & $1^1$A$^{\prime\prime}$ ($n\rightarrow\pi^*$; V) & 3.94 & 3.74 &  3.59 & 4.31 & 4.27 \\
\\
acetylene & $1^1\Delta_u (\pi\rightarrow\pi^*$; V) & 6.69 & 7.72 & 7.76 &7.10 & 7.2 \\
\\
ammonia & 2$^1$A$_1$(n$\rightarrow$3s; R) & 6.44 & 6.40 & 6.37 & 6.66 & 6.38 \\
\\
carbon monoxide & $1^1\Pi$(n$\rightarrow\pi^*$; V) & 7.48 & 7.96  & 9.36 & 8.48 & 8.51 \\
\\
diazomethane & 1$^1$A$_2$($\pi\rightarrow\pi^*$; V) & 2.94 & 2.56 & 2.14 & 3.13 & 3.14 \\
\\
\\
ethylene & 1$^1$B$_{3u}$(n$\rightarrow$3s; R) & 7.14 & 7.18 & 7.2 & 7.43 & 7.11 \\
  & 1$^1$B$_{1u}$($\pi\rightarrow\pi^*$; V) & 6.72 & 7.17 & 7.64 & 7.92 & 7.6 \\
\\
formaldehyde & 1$^1$B$_2$(n$\rightarrow$3s; R) & 6.89 & 7.10 & 7.30 & 7.11 \\
\\
formamide & 1$^1$A"(n$\rightarrow\pi^*$; V) & 5.38 & 5.17 & 5.01 & 5.63 & 5.8 \\
\\
hydrogen sulfide & 1$^1$A$_2$(n$\rightarrow$4p; R) & 5.63 & 5.63 & 5.55 & 6.10 &  \\
\\
ketene & 1$^1$B$_1$($\pi\rightarrow$3s; R) & 5.87 & 5.97 & 6.11 & 6.06 & 5.86 \\
\\
methanimine & 1$^1$A"(n$\rightarrow\pi^*$; V) & 4.65 & 4.77 & 4.89 & 5.21 & \\
\\
thioformaldehyde & 1$^1$A$_2$(n$\rightarrow\pi^*$; V) & 1.91 & 1.74 & 1.57 & 2.20 & 2.03 \\
  & 1$^1$B$_2$(n$\rightarrow$4s; R) & 5.64 & 5.72  & 5.77 & 5.99 & 5.85 \\
  & 2$^1$A$_1$($\pi\rightarrow\pi^*$; V) & 5.36 & 6.02 &  6.66 &  6.34 & 6.2 \\
\\
water molecule& 1$^1$B$_1$(n$\rightarrow$3s; R) & 7.46 & 7.46 & 7.41 & 7.70 & 7.41 \\
  & 1$^1$A$_2$(n$\rightarrow$3p; R) & 8.91 & 9.02 & 9.11 & 9.47 & 9.2 \\
  & 2$^1$A$_1$(n$\rightarrow$3s; R) & 9.73 & 9.71 & 9.69 & 9.97 & 9.67 \\
\\
\hline
\rowcolor{Gray}
& ME (TBE) & -0.46 & -0.33 & -0.21 & &\\
& ME (EXP) & -0.30 & -0.17& -0.06 & &\\
\rowcolor{Gray}
& MAE (TBE) & 0.46 & 0.40& 0.44 & & \\
& MAE (EXP) & 0.33 & 0.27& 0.36 & &\\
\rowcolor{Gray}
& RMSE (TBE) & 0.54 & 0.43 & 0.51 & & \\
& RMSE (EXP) & 0.46 & 0.35 & 0.49 & & \\
\end{tabular}

  \textsuperscript{\emph{a}}Corrected theoretical best estimates as given in Ref.~\cite{Loos2018}. \ 
  \textsuperscript{\emph{b}}Experimental values listed in Ref.~\cite{Loos2018} (see references therein).
\end{table}

\begin{table}[H]
\footnotesize
  \caption{
  One-electron self-interaction energy in eV calculated for the ground- and excited-state optimal orbitals of the water molecule. 
  The excited-state 3$s$ orbital has smaller classical self-Coulomb energy than the valence ground-state (GS) orbitals. 
  However, the estimate of the total self-interaction energy is of similar magnitude for all the orbitals.
  }
  \label{tbl: si_water}
  \begin{tabular}{ l  | c  c  c}
  orbital &  Coulomb & XC &  Coulomb + XC  \\ 
  \hline
  GS$_1$\textsuperscript{\emph{a}} & 10.06 & -10.46 & -0.40 \\
  GS$_2$\textsuperscript{\emph{a}} & 10.98 & -11.23 &  -0.26 \\ 
  $3s$ orbital\textsuperscript{\emph{b}} & 3.34 & -3.70 & -0.36\\
  \hline
\end{tabular}

\textsuperscript{\emph{a}}Two orbitals from a ground-state calculation with PBE-SIC/2 (see Figure~S5 in the Supplementary Information),
giving different SIC estimates. \textsuperscript{\emph{b}}Orbital obtained from a DO-MOM excited-state calculation.
\end{table}

\begin{figure}[H] 
    \centering
    \includegraphics[width=1\textwidth]{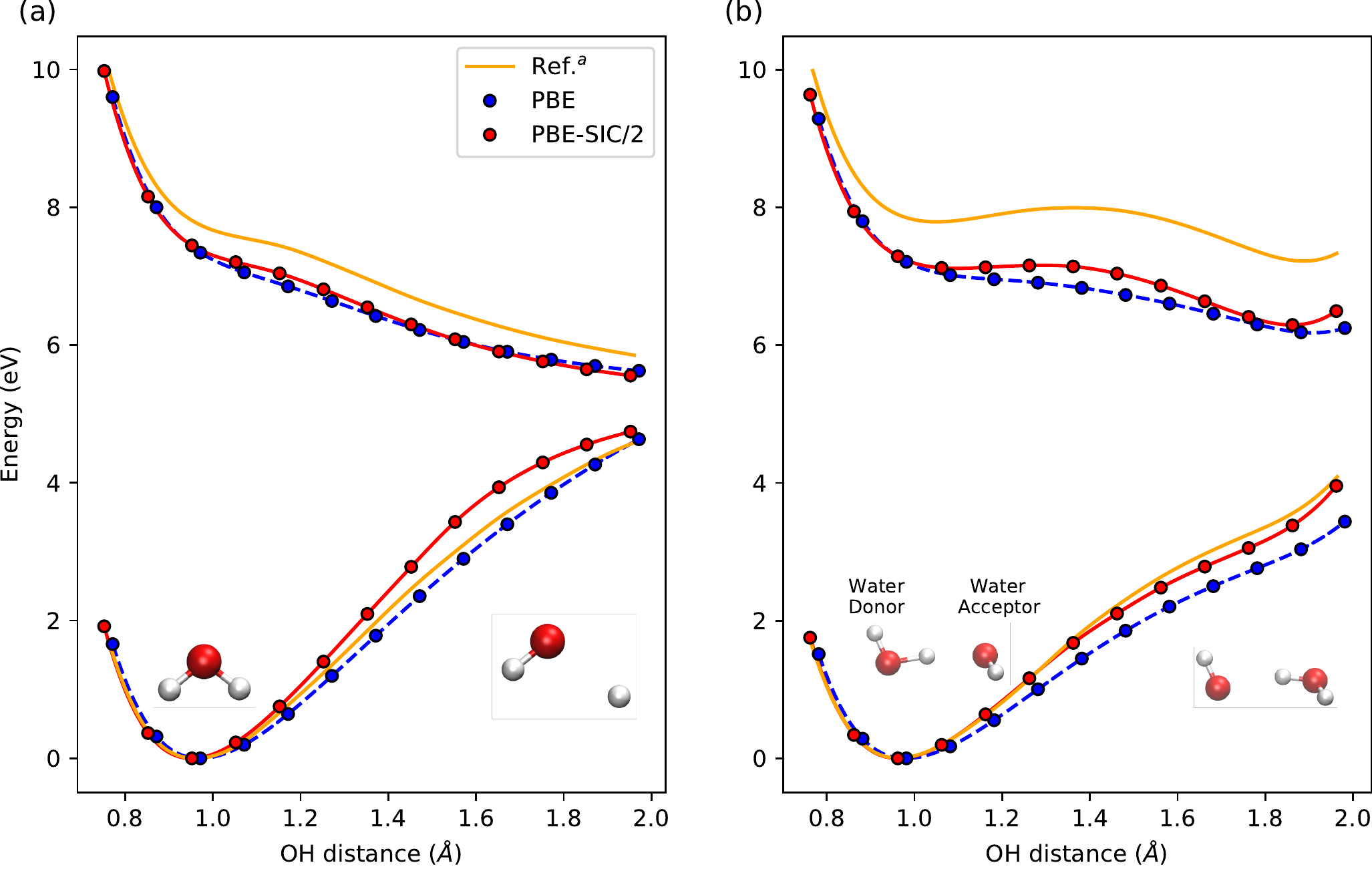}
    \caption{Potential energy curves calculated for ground and lowest excited singlet states
    along the hydrogen-bonded O-H bond with all other molecular geometry parameters fixed. (a) Water monomer, (b) Water dimer.
    The insets show the atomic configurations 
     at the energy minimum and the endpoint of the scan in the O-H bond length.
    {}\textsuperscript{\emph{a}}CR-EOMCCD(T),ID/aug-cc-pVTZ calculations from Ref.~\cite{Chipman2006} } 
    \label{fig:Water}
\end{figure}

In order to further investigate the effect of PZ-SIC on excited states, 
the change in energy as an O-H bond is stretched in a water molecule and a dimer of water molecules is calculated for
the ground and lowest singlet excited states. 
First, the ground-state geometry is optimized until the maximum of the force on the atoms has magnitude below 0.01 eV/\AA. 
Then, the hydrogen-bonded O-H bond is stretched by changing the position of the hydrogen atom in increments of 0.1 {\AA}, 
while keeping the positions of all other atoms frozen. The results of these calculations with both PBE and PBE-SIC/2 are 
presented in Fig.~\ref{fig:Water} and compared with the results of CR-EOMCCSD(T),ID/aug-cc-pVTZ calculations from Ref.~\cite{Chipman2006}. 
The energy curves obtained with PBE and PBE-SIC/2 lie close to each other. 
However, the PBE-SIC/2 energy curve reproduces better the S-shape of the high level reference curves for both the monomer and the dimer. 
For the water dimer in the lowest excited state, PBE-SIC/2 reproduces the local minimum at short bond length 
while PBE predicts a barrierless path towards the second constrained energy minimum near 1.8 {\AA} O-H distance. 

\section{Discussion and Conclusions}\label{sec: discandconc}
The RSG or PW representations have an advantage over LCAO in that full basis set limit can be reached by systematically changing only a single parameter. 
However, such calculations involve larger computational cost than LCAO and are limited to the frozen core approximation. 
A useful strategy for reducing the computational cost is to obtain initial orbitals from an LCAO calculation of the excited state 
and then switch to RSG or PW mode. All three types of representations can be available in the same software, as is the case with our implementation
in the GPAW software,
making such a hybrid approach relatively straightforward.

The search of saddle points on multidimensional electronic energy surfaces is a considerably more difficult problem than minimization due to the fact that the energy must be maximized along a few degrees of freedom that are not known {\it a priori}. Even if the order of a saddle point is known, the success of the extremization of the energy depends on how close the initial guess for the orbitals is to the target solution. The ground-state orbitals can represent a good enough initial guess for the iterative convergence on an excited state, but in some cases another excited-state solution provides a better choice. 
This can be especially important when the lowest energy excited state does not correspond to a HOMO-LUMO excitation. 
For example, in the water dimer the lowest excited state corresponds to an electron-hole pair localized on the hydrogen bond donor molecule but the HOMO-LUMO excitation corresponds to the transfer of an electron from the donor to the acceptor molecule with the hole localized on the donor and the excited electron localized on the acceptor.  In the calculations of the potential energy curve for the lowest excited state in the water dimer presented in Sec.~\ref{sec: PZ-SIC}, a two-step procedure was used: First, the HOMO-LUMO excitation was computed and the obtained solution then used as an initial guess in the calculation on the lowest excited state.
 
In conclusion, we have developed a formulation of the DO-MOM algorithm for density functional calculations of excited states that can be applied to
RSG and PW representations,
where it is not possible to include all
virtual orbitals, unlike with LCAO basis sets. This implementation of 
the DO-MOM algorithm is found to be robust and able to converge on
excited states that are challenging for commonly used SCF-MOM algorithms. The importance of complex-valued orbitals in calculations of excited states with a KS-DFT functional is also demonstrated. In the case of the lowest excited state of CO, it was shown that real-valued orbitals break the uniaxial symmetry of the electron density, while complex-valued orbitals restore the symmetry and
correspond to higher energy solutions. The fact that complex orbitals provide a better description of the electronic system is in par with findings from ground-state calculations~\cite{Klupfel2011,Klupfel2012mol,Lehtola2016b,Small2015,Lee2019}.

 The application of DO-MOM in calculations of the excited states of the [Fe(bmip)$_2$]$^{2+}$ complex shows that: (1) the approach is robust enough to allow the calculation of several excited states that are close in energy, including regions where the potential energy curves cross; and (2) the predicted excited-state
 properties are in good agreement with experimental values.
 As for any single-determinant method, the quality of the excited states obtained with DO-MOM is expected to degrade when the states have a multi-determinantal character and this can represent a limitation for the application of the methodology to dynamics simulations. 
 However, when explicit solvent effects are included in the simulations, the symmetry breaking induced by the solvent can lift degeneracy, thereby reducing the presence of multi-determinantal states.
So far, excited-state dynamics simulations of systems as large as transition metal complexes using a variational eDFT approach have been limited to a single Born-Oppenheimer surface \cite{Levi2019, Levi2018, Dohn2014}.
The present results represent a preliminary indication that the DO-MOM method is viable for nonadiabatic molecular dynamics simulations including multiple excited states. Current efforts in this direction include evaluating nonadiabatic couplings between the excited states obtained variationally, using numerical or analytical approaches such as the one recently presented in Ref.\cite{Ramos2021}. A simulation of the excited-state relaxation of the [Fe(bmip)$_2$]$^{2+}$ complex starting from an excitation to the bright MLCT state could help explain the experimentally observed ultrafast $^3$MLCT/$^3$MC branching that affects its performance 
as a photosensitizer.

 An assessment of self-interaction correction, the PZ-SIC, applied to the PBE functional has been performed for excited states. Variational calculations with SIC generally improve the description of Rydberg states, which is in agreement with previous studies where linear-expansion $\Delta$SCF~\cite{Gavnholt2008} calculations with the PBE functional estimated on SIC orbitals were performed~\cite{Gudmundsdottir2013,Gudmundsdottir2014}.
 For valence triplet states, the correction tends to reduce the excitation energy when the SIC in the ground state is larger than that in the excited state. At the same time, SIC sometimes might have a small effect due to a cancellation of the correction to the energy of the ground and excited states, which
 may be attributed to the one-electron nature of this orbital-by-orbital estimate of the correction.
 Both the PBE value and the corrected value typically underestimate the excitation energy. 
 Even though the effect of SIC on the excitation energy can be small, it can lead to improved shape of the potential energy surface as is seen 
 for the lowest excited state of the water monomer and dimer. There, the dependence of the excited-state energy on the O-H bond length agrees with 
 results of high-level, CR-EOMCCD(T),ID/aug-cc-pVTZ calculations~\cite{Chipman2006} better than the results obtained with the PBE functional. 
 Further improvement in excited-state energy calculations 
 may require going beyond the self-interaction correction based on orbital densities, for example by including the complementary density in the error estimate~\cite{Lundin2001}. 

An important advantage of the RSG and PW implementation presented here is the possibility of doing calculations for extended systems that are
subject to periodic boundary conditions. This includes, for example, excited states of defects in crystals and solvent effects on 
electronic excitations of molecular complexes. This will be the topic of future studies with the DO-MOM method.


\section*{Supporting Information}
The Supporting Information includes the following: Isosurfaces of the ground-state molecular orbitals of the [Fe(bmip)$_2$]$^{2+}$ complex. A comparison between the potential energy curves of the MC$^3$ state of [Fe(bmip)$_2$]$^{2+}$ as calculated with DO-MOM and TDDFT. An analysis of the excitation energies for the molecules in Section \ref{sec: PZ-SIC} as obtained from the mixed-spin determinant and with PBE0 and PBE50, as well as PBE-SIC/2 restricted to real orbitals. Isosurfaces of the ground-state PBE-SIC complex orbitals of the water molecule. 


\appendix
\section*{Appendix}
\renewcommand{\thesubsection}{\Alph{subsection}}

\renewcommand{\theequation}{\thesubsection.\arabic{equation}}
\numberwithin{equation}{subsection}

\subsection{Inner-Loop Minimizaion for ODD functional}\label{appendix1}
The unitary matrix that minimizes an ODD functional, 
such as the PZ-SIC 
among occupied orbitals is found using an exponential transformation:
\begin{equation}
\ket{\psi_i} = \sum_{ij} (e^{-A})_{ij} \ket{\phi_j} 
\end{equation}
where $A$ is a skew-hermitian matrix, $A^{\dagger} = -A$. Let 
\begin{equation}
    \lambda_{ij} = \Braket{\phi_i| \hat h_{j} + \hat v_j |\phi_j}
\end{equation}
with $\hat h_j$ defined in Eq.~\eqref{eq: gradE} and 
\begin{equation}
    \hat v_j =  -  \frac{\delta \left( \cU[\rho_{j}] +  \cE_{xc}[\rho_{j}, 0]\right)}{\delta \rho_{j}}
\end{equation}
The gradient of the energy with respect to matrix elements of A is:
\begin{equation}\label{eq: dedA}
    \frac{\partial \cE^{ODD}}{\partial A_{ij}} = \lambda_{ij} - \lambda_{ji}^{*} + o(\|A\|)
\end{equation}
This expression must be zero at the minimum and for equally occupied orbitals this leads to the Pederson (or localization) conditions~\cite{Pederson1984,Pederson1985a}:
\begin{equation}
     \Braket{\phi_i| \hat v_j - \hat v_i |\phi_j} = 0
\end{equation}

The L-BFGS algorithm with inexact line search is used to find the optimal matrix $A$ and, thereby, the
optimal unitary matrix $O=e^{-A}$. 
Details of the implementation of the exponential transformation are given in Ref.~\cite{Ivanov2020}.

\subsection{Maximum Overlap Method within the Projector-Augmented Wave Approach}\label{appendix2}
Let $\{\ket{\psi_n}\}$ be the $N$ orbitals used as initial guess for an excited-state calculation and $\{\ket{\psi^{(k)}_m}\}$ the $M$ orbitals at the $k$\textsuperscript{th} iteration of the optimization. Within the PAW approach, the elements of the overlap matrix between orbitals $\{\ket{\psi_n}\}$ and $\{\ket{\psi^{(k)}_m}\}$ are~\cite{PAW_Blochl:1994}:

\begin{equation}
    S_{nm}^{(k)} = \Braket{\psi_n | \psi_{m}^{(k)}} = 
    \Braket{\tilde{\psi}_n | \tilde{\psi}_{m}^{(k)}} + \sum_{a, i_1, i_2} \Braket{\tilde{\phi}_n | \tilde{p}_{i_1}^{a}} \left( 
     \Braket{ \phi_{i_1}^{a} | \phi_{i_2}^{a}}
     -
     \Braket{ \tilde{\phi}_{i_1}^{a} | \tilde{\phi}_{i_2}^{a}}
    \right) \Braket{ \tilde{p}_{i_2}^{a} | \tilde{\psi}_{m}^{(k)}}
\end{equation}
where $\ket{\tilde{\psi}_n}$ are pseudo orbitals, $\ket{\tilde{p}_{i_1}^{a}}$ ($\ket{\tilde{p}_{i_2}^{a}}$) are projector functions localized on atom $a$, while $\ket{\phi_{i_1}^{a}}$ ($\ket{\phi_{i_2}^{a}}$) and $\ket{\tilde{\phi}_{i_1}^{a}}$ ($\ket{\tilde{\phi}_{i_2}^{a}}$) are partial and pseudo partial waves localized on atom $a$, respectively. 
After computing the overlap matrix at iteration $k$, the occupation numbers $\{f_m\}$ of the orbitals $\{\ket{\psi^{(k)}_m}\}$ are chosen as following. An occupation number of 1 is given to the first $N$ orbitals with the highest numerical weights, evaluated from the projection onto the manifold $\{\ket{\psi_n}\}$: 
\begin{equation}
    \omega_{m}^{(k)} = \left(\sum_{n=1}^{N}  |S_{nm}^{(k)}|^{2} \right)^{1/2}
\end{equation}
The remaining $M-N$ orbitals are left unoccupied (occupation number of 0).


%
%
%


\begin{acknowledgement}
This work was supported by the University of Iceland Research Fund and the Icelandic Research Fund.
AVI is supported by a doctoral fellowship from the University of Iceland.
GL thanks the Icelandic Research Fund for financial support under grant number 196070.
The authors are grateful to Mátyás Pápai for discussions on the Fe photosensitizer complex.
The calculations were carried out at the Icelandic Research High Performance Computing (IRHPC) 
facility at the University of Iceland.
\end{acknowledgement}

\bibliography{main}

\end{document}


\tableofcontents

\clearpage
\section{Ground-state molecular orbitals of [Fe(bmip)$_2$]$^{2+}$}
Isosurfaces showing the molecular orbitals of the [Fe(bmip)$_2$]$^{2+}$ complex involved in the electronic transitions investigated in the present work. 
The isofurfaces are represented with gray and orange colors and correspond to isovalues of $\pm0.16$ {\AA}$^{-3/2}$.
\begin{figure}[H]
    \centering
    \includegraphics[width=0.7\textwidth]{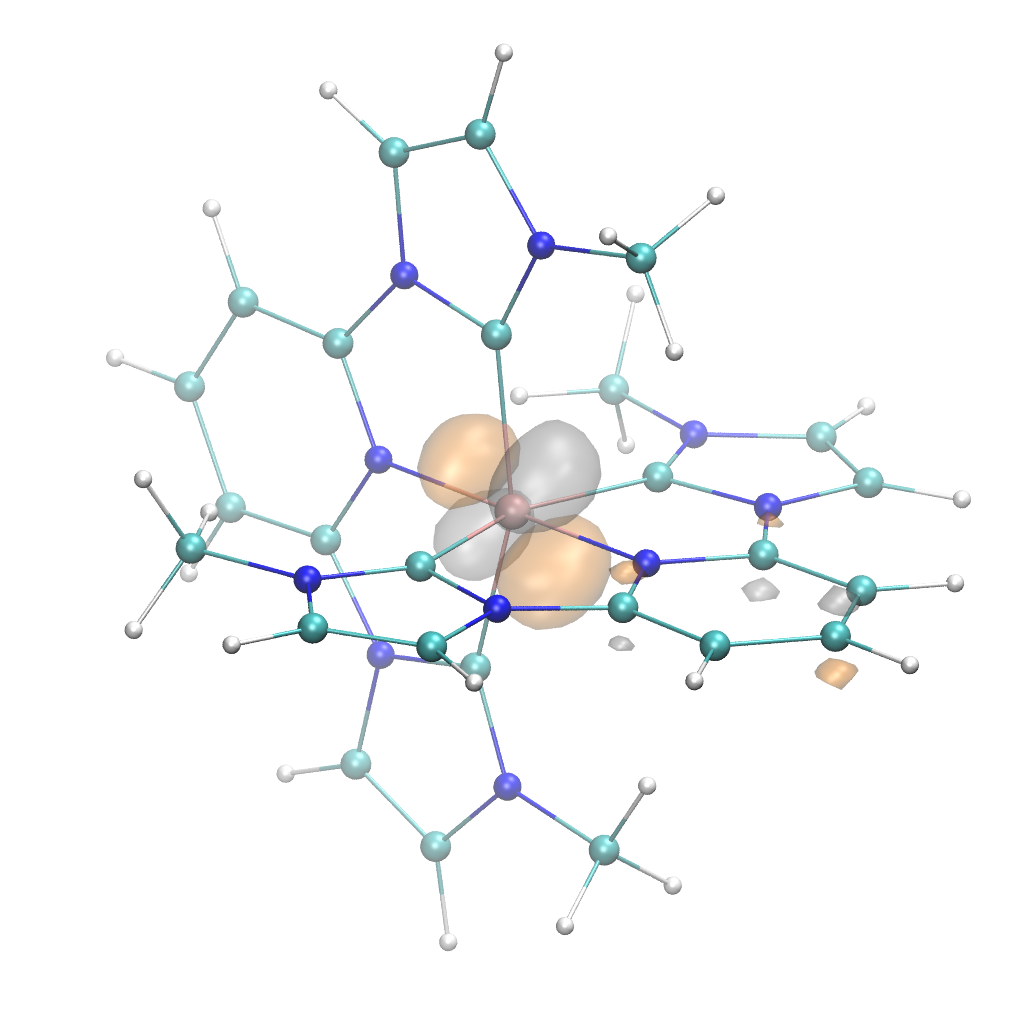}
    \caption{HOMO-1} 
    \label{fig: HOMO-1}
\end{figure}

\begin{figure}[H]
    \centering
    \includegraphics[width=0.7\textwidth]{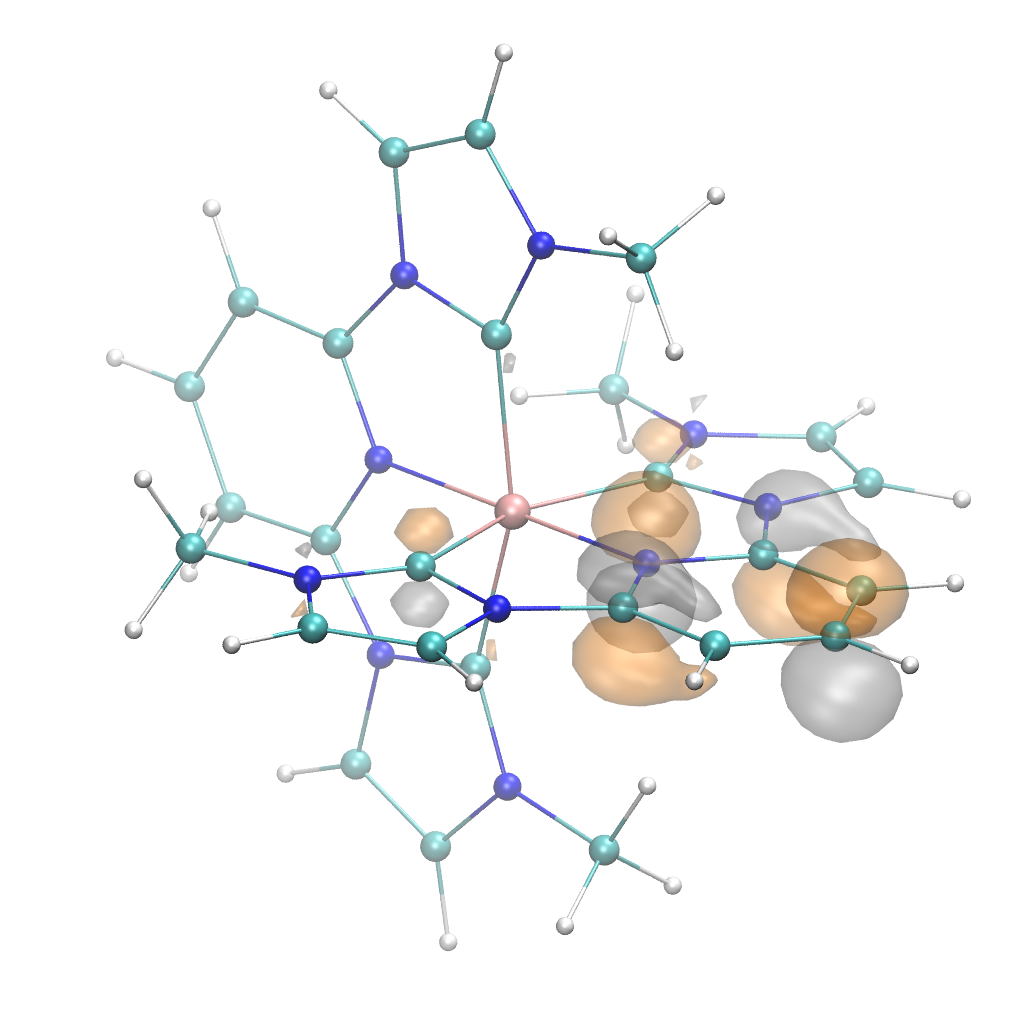}
    \caption{LUMO} 
    \label{fig: LUMO}
\end{figure}

\begin{figure}[H]
    \centering
    \includegraphics[width=0.7\textwidth]{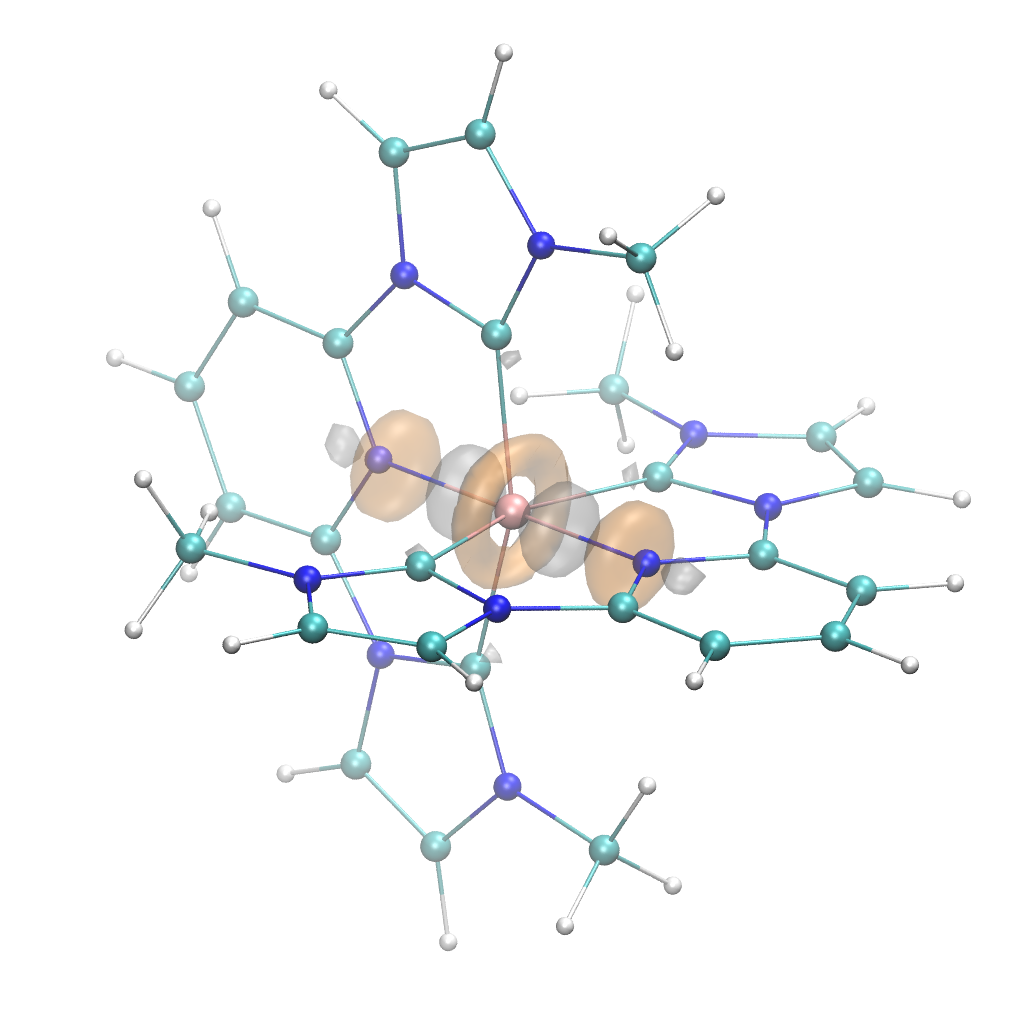}
    \caption{LUMO+4} 
    \label{fig: LUMO+4}
\end{figure}

\clearpage

\section{Curvature of the $^3$MC potential energy curve of [Fe(bmip)$_2$]$^{2+}$}

\begin{figure}[H]
    \centering
    \includegraphics[width=1\textwidth]{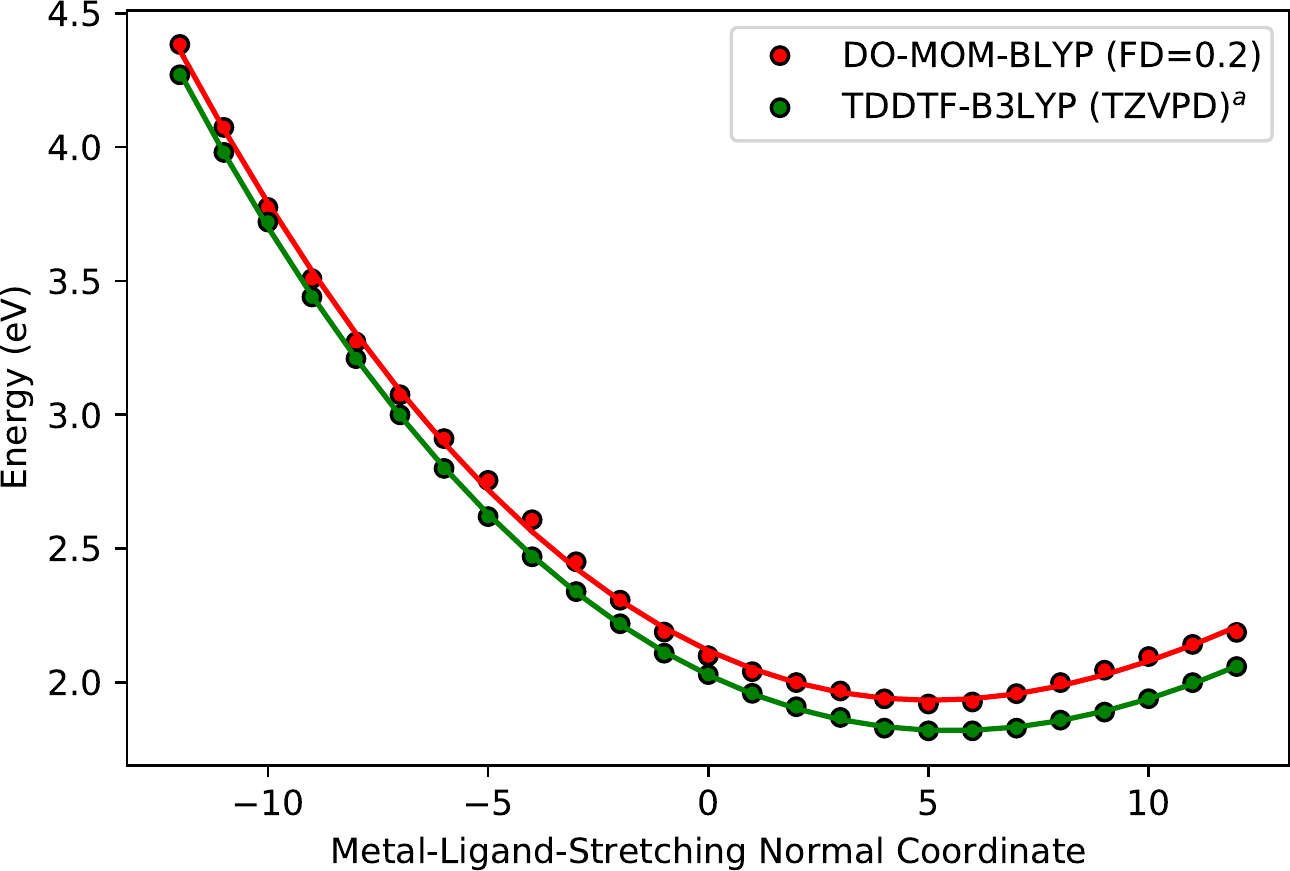}
    \caption{Potential energy curves along a metal-ligand normal coordinate (the breathing mode $Q_6$ as defined in Ref.~\cite{Papai2019a}) of the lowest $^3$MC state. The DO-MOM (BLYP) calculations were performed using a finite-difference real-space basis set as described in the main text of the article.
    \textsuperscript{a}Calculations from Ref.~\cite{Papai2019a}.} 
    \label{fig: MC3}
\end{figure}

The potential energy curves of the lowest $^3$MC state of [Fe(bmip)$_2$]$^{2+}$ calculated with 
DO-MOM (BLYP) in the present work and linear response TDDFT (B3LYP*) from Ref.~\cite{Papai2019a}
are fitted using a fourth order polynomial:
\begin{equation}
E(Q) = -d_e + \frac{k}{2} (Q - r_e)^2 - \alpha(Q - r_e)^3 + \beta (Q - r_e)^4,
\end{equation}
where $k$ is the curvature. Below are the results of the fitting of the curve calculated with 
DO-MOM (BLYP):
\begin{align*}
    &d_e = -1.934 \\
    &k = 1.332\times10^{-2} \\ 
    &r_e = 5.060 \\ 
    &\alpha = 1.265\times10^{-4} \\
    &\beta = -1.652\times10^{-6} 
\end{align*}
and of the curve calculated with TDDFT (B3LYP*):
\begin{align*}
    &d_e = -1.821 \\
    &k = 1.263\times10^{-2} \\ 
    &r_e = 5.514 \\ 
    &\alpha = 9.256\times10^{-4} \\
    &\beta = 2.461\times10^{-7} 
\end{align*}
The fitted curves are shown in Fig.~\ref{fig: MC3} together with the points obtained from the DO-MOM (BLYP) and TDDFT (B3LYP*) calculations. The curvatures of the fitted curves differ by around 5\%. The oscillation period estimated on the basis of TDDFT is 285 fs~\cite{Papai2019a}, and thus, we obtain an estimation of an oscillation period on the basis of eDFT as \[ T_{eDFT} = T_{TDDFT} \sqrt{\frac{k_{TDDFT}}{k_{eDFT}}} \approx 280\, (\text{fs}) \]

\clearpage

\section{Excitation Energy obtained from the Mixed-Spin Determinant}

\begin{table}[H]
\scriptsize
 \caption{Mixed-Spin States}
 \label{tbl: singlets}
 \begin{tabular}{ l  c  c  c  c  c c}
    \hline
    species & excitation & PBE\textsuperscript{\emph{a}} & SIC/2\textsuperscript{\emph{b}} & SIC\textsuperscript{\emph{b}} & TBE\textsuperscript{\emph{c}} & EXP\textsuperscript{\emph{d}} \\
    \hline
acetaldehyde & $1^1$A$^{\prime\prime}$ ($n\rightarrow\pi^*$; V) & 3.79 & 3.74 & 3.69 &4.31 & 4.27 \\
\\
acetylene & $1^1\Delta_u (\pi\rightarrow\pi^*$; V) & 6.51 & 6.81 & 6.90 &7.10 & 7.2 \\
\\
ammonia & 2$^1$A$_1$(n$\rightarrow$3s; R) & 6.30 & 6.25 & 6.23 & 6.66 & 6.38 \\
\\
carbon monoxide & $1^1\Pi$(n$\rightarrow\pi^*$; V) & 6.70 & 6.90 & 7.51 & 8.48 & 8.51 \\
\\
diazomethane & 1$^1$A$_2$($\pi\rightarrow\pi^*$; V) & 2.85 & 2.47 & 2.01 & 3.13 & 3.14 \\
\\
ethylene & 1$^1$B$_{3u}$(n$\rightarrow$3s; R) & 7.07 & 7.12 & 7.14 & 7.43 & 7.11 \\
  & 1$^1$B$_{1u}$($\pi\rightarrow\pi^*$; V) & 5.59 & 5.90 & 6.20 & 7.92 & 7.6 \\
\\
formaldehyde & 1$^1$B$_2$(n$\rightarrow$3s; R) & 6.79 & 7.04 & 7.16 & 7.30 & 7.11 \\
\\
formamide & 1$^1$A"(n$\rightarrow\pi^*$; V) & 5.26 & 5.20 & 5.14 & 5.63 & 5.8 \\
\\
hydrogen sulfide & 1$^1$A$_2$(n$\rightarrow$4p; R) & 5.51 & 5.53 & 5.49& 6.10 &         \\
\\
ketene & 1$^1$B$_1$($\pi\rightarrow$3s; R) & 5.75 & 5.87 & 5.95 & 6.06 & 5.86 \\
\\
methanimine & 1$^1$A"(n$\rightarrow\pi^*$; V) & 4.43 & 4.56 & 4.65 & 5.21 &         \\
\\
thioformaldehyde & 1$^1$A$_2$(n$\rightarrow\pi^*$; V) & 1.81 & 1.78 & 1.72 & 2.20 & 2.03 \\
  & 1$^1$B$_2$(n$\rightarrow$4s; R) & 5.47 & 5.63 & 5.72 & 5.99 & 5.85 \\
  & 2$^1$A$_1$($\pi\rightarrow\pi^*$; V) & 4.36 & 4.67 & 4.97 & 6.34 & 6.2 \\
\\
water & 1$^1$B$_1$(n$\rightarrow$3s; R) & 7.28 & 7.27 &  7.24 & 7.70 & 7.41 \\
  & 1$^1$A$_2$(n$\rightarrow$3p; R) & 8.83 & 8.94 & 9.04 & 9.47 & 9.2 \\
  & 2$^1$A$_1$(n$\rightarrow$3s; R) & 9.50 & 9.48 & 9.46 & 9.97 & 9.67 \\
\\
\hline
\rowcolor{Gray}
& ME (TBE) & -0.73 & -0.66 & -0.59 & &\\
& ME (EXP) & -0.59 & -0.52 & -0.49 & &\\
\rowcolor{Gray}
& MAE (TBE) & 0.73 & 0.66 & 0.59 & &\\
& MAE (EXP) & 0.59 & 0.52 & 0.50 & &\\
\rowcolor{Gray}
& RMSE (TBE) & 0.95 & 0.83 & 0.73 & &\\
& RMSE (EXP) & 0.88 & 0.76 & 0.68 & &\\
\end{tabular}

  \textsuperscript{\emph{a}} Employing real-valued orbitals;
  \textsuperscript{\emph{b}} Initial guess for the wave functions is PBE real-valued orbitals followed by complex Rudenberg-Edmiston localization;
  \textsuperscript{\emph{c}} Corrected theoretical best estimates as given in Ref.~\cite{Loos2018};
  \textsuperscript{\emph{d}} Listed in Ref.~\cite{Loos2018} (see references therein);

\end{table}

\clearpage
\section{Extended Data For Excitation Energy Including Hybrid Functionals}
Calculations with hybrid functionals are performed using the real space grid (RSG) implementation in the GPAW software~\cite{GPAW2} with grid spacing and simulation box as described in the main text. Some excitation energies were first compared with all-electron calculation using the ORCA software\cite{neese2012,neese2018}, in order to test the validity of the RSG hybrid functional calculations in GPAW. The ORCA calculations made use of a Def2-QZVVPD basis set, no resolution of identity for the calculation of exact exchange, and fifth grid level. For the lowest triplet excitations in diazomethane and thioformaldehyde, the excitation energies given by PBE0 and calculated with ORCA are found to be -2.482 eV and -1.596 eV, respectively, and those calculated with PBE0 and the GPAW software are found to be -2.482 and -1.594 eV, respectively. This indicates that the frozen core approximation and approximations due to the PAW approach have little affect on the low lying excitation energies. 

\begin{table}[H]
\scriptsize
  \caption{
Energy of excitations to triplet states calculated with the DO-MOM method and comparison with theoretical
best estimates as well as experimental values. The calculations make use of a generalized gradient
approximation Kohn-Sham functional (PBE), with scaled self-interaction correction (SIC/2),
with full self-interaction correction (SIC) as well as two hybrid functionals PBE0 with 25\% of exact exchange and PBE50 with 50\% of exact exchange. The mean error (ME), mean absolute error (MAE) and
root mean square error (RMSE) are given with respect to theoretical best estimates and with 
respect to experimental values at the bottom of the table.
  }
  \label{tbl: triplets}
  \begin{tabular}{ l  c c c c c c c c }
    \hline
    molecule & excitation &  PBE & PBE0 & PBE50 & SIC/2 & SIC &  TBE\textsuperscript{\emph{a}} & EXP\textsuperscript{\emph{b}} \\
    \hline
acetaldehyde & 1$^3$A$^{\prime\prime}$ ($n\rightarrow\pi^*$; V) & 3.65 & 3.58 & 3.45 & 3.75 & 3.79 & 3.98 & 3.97 \\
\\
acetylene & $1^3\Delta_u$($\pi\rightarrow\pi^*$; V) & 6.33 & 6.2 & 6.07 & 5.89 & 6.04 & 6.40 & 6.0 \\
\\
ammonia & 2$^3$A$_1$(n$\rightarrow$3s; R) & 6.16 & 6.14 & 6.12 & 6.10 & 6.06 & 6.37 & 6.02 \\
\\
carbon monoxide & $1^3\Pi$(n$\rightarrow\pi^*$; V) & 5.91 & 5.88 & 5.82 & 5.66 & 6.28 & 6.32 \\
\\
diazomethane & 1$^3$A$_2$($\pi\rightarrow\pi^*$; V) & 2.76 & 2.48 & 2.18 & 2.38 & 1.88 & 2.80 &         \\
\\
\\
ethylene & 1$^3$B$_{3u}$(n$\rightarrow$3s; R) & 7.01 & 6.95 & 6.89 & 7.05 & 7.07 & 7.28 & 6.98 \\
  & 1$^3$B$_{1u}$($\pi\rightarrow\pi^*$; V) & 4.46& 4.25 & 4.04 & 4.63 & 4.75 & 4.54 & 4.36 \\
\\
formaldehyde & 1$^3$B$_2$(n$\rightarrow$3s; R) & 6.69& 6.83 & 6.91 & 6.99 & 7.12 & 7.14 & 6.83 \\
\\
formamide & 1$^3$A"(n$\rightarrow\pi^*$; V) & 5.14& 5.0 & 4.82 &5.23 & 5.27 & 5.37 & 5.2 \\
\\
hydrogen sulfide & 1$^3$A$_2$(n$\rightarrow$4p; R) & 5.39& 5.45 & 5.49 & 5.44 & 5.43 & 5.74 & 5.8 \\
\\
ketene & 1$^3$B$_1$($\pi\rightarrow$3s; R) & 5.64& 5.76 & 5.87 & 5.77 & 5.79 & 5.85 & 5.8 \\
\\
methanimine & 1$^3$A"(n$\rightarrow\pi^*$; V) & 4.20& 4.17 & 4.10 & 4.35 & 4.41 & 4.64 &         \\
\\
thioformaldehyde & 1$^3$A$_2$(n$\rightarrow\pi^*$; V) & 1.71& 1.59 & 1.46 & 1.81 & 1.88 & 1.94 &         \\
  & 1$^3$B$_2$(n$\rightarrow$4s; R) & 5.31 & 5.47 & 5.61 &5.54 &  5.67 & 5.76 &         \\
  & 2$^3$A$_1$($\pi\rightarrow\pi^*$; V) & 3.36 & 3.09 & 2.81 & 3.33 & 3.28 & 3.44 & 3.28 \\
\\
water molecule& 1$^3$B$_1$(n$\rightarrow$3s; R) & 7.10 & 7.02 & 6.94 & 7.09 & 7.08 & 7.33 & 7.2 \\
  & 1$^3$A$_2$(n$\rightarrow$3p; R) & 8.75 & 8.74 & 8.72 & 8.87 & 8.97 & 9.30 & 8.9 \\
  & 2$^3$A$_1$(n$\rightarrow$3s; R) & 9.28& 9.24 & 9.2 & 9.25 & 9.23 & 9.59 & 9.46 \\
\\
\hline
\rowcolor{Gray}
& ME (TBE) & -0.27 & -0.33 & -0.41 & -0.25& -0.25& & \\
& ME (EXP) & -0.09 & -0.14 & -0.21 & -0.06 &-0.04& &  \\
\rowcolor{Gray}
& MAE (TBE) & 0.27 & 0.33 & 0.41 & 0.26 & 0.27 & &  \\
& MAE (EXP) & 0.19 & 0.19 & 0.26 & 0.16 & 0.18 & &  \\
\rowcolor{Gray}
& RMSE (TBE) & 0.31 & 0.35 & 0.44 & 0.29 & 0.34 & & \\
& RMSE (EXP) & 0.22 & 0.23 & 0.30 & 0.21 & 0.26 & & \\
\end{tabular}

  \textsuperscript{\emph{a}}Theoretical best estimates as given in Ref.~\cite{Loos2018}. \ 
  \textsuperscript{\emph{b}}Experimental values listed in Ref.~\cite{Loos2018} (see references therein).
\end{table}


\begin{table}[H]
\scriptsize
  \caption{
Energy of excitations to singlet states (spin purified) calculated with the DO-MOM method and comparison with theoretical
best estimates as well as experimental values. The calculations make use of a generalized gradient
approximation Kohn-Sham functional (PBE), with scaled self-interaction correction (SIC/2),
with full self-interaction correction (SIC) as well as two hybrid functionals PBE0 with 25\% of exact exchange and PBE50 with 50\% of exact exchange. The mean error (ME), mean absolute error (MAE) and
root mean square error (RMSE) are given with respect to theoretical best estimates and with 
respect to experimental values at the bottom of the table.
  }
  \label{tbl: pure-singlets}
  \begin{tabular}{ l  c  c c c  c  c c c}
    \hline
    molecule & excitation & PBE & PBE0 & PBE50 & SIC/2 &
    SIC &
    TBE\textsuperscript{\emph{a}} & EXP\textsuperscript{\emph{b}} \\
    \hline
acetaldehyde & $1^1$A$^{\prime\prime}$ ($n\rightarrow\pi^*$; V) & 3.94 & 3.85 & 3.70 & 3.74 &  3.59 & 4.31 & 4.27 \\
\\
acetylene & $1^1\Delta_u (\pi\rightarrow\pi^*$; V) & 6.69 & 6.59 & 6.47 & 7.72 & 7.76 &7.10 & 7.2 \\
\\
ammonia & 2$^1$A$_1$(n$\rightarrow$3s; R) & 6.44 & 6.42 & 6.41 & 6.40 & 6.37 & 6.66 & 6.38 \\
\\
carbon monoxide & $1^1\Pi$(n$\rightarrow\pi^*$; V) & 7.48 & 7.83 & 8.17 & 7.96  & 9.36 & 8.48 & 8.51 \\
\\
diazomethane & 1$^1$A$_2$($\pi\rightarrow\pi^*$; V) & 2.94 & 2.66 & 2.36 & 2.56 & 2.14 & 3.13 & 3.14 \\
\\
\\
ethylene & 1$^1$B$_{3u}$(n$\rightarrow$3s; R) & 7.14 & 7.41 & 6.99 & 7.18 & 7.2 & 7.43 & 7.11 \\
  & 1$^1$B$_{1u}$($\pi\rightarrow\pi^*$; V) & 6.72 & 7.07 & 8.09 & 7.17 & 7.64 & 7.92 & 7.6 \\
\\
formaldehyde & 1$^1$B$_2$(n$\rightarrow$3s; R) & 6.89 & 6.98 & 7.01 & 7.10 & 7.30 & 7.11 \\
\\
formamide & 1$^1$A"(n$\rightarrow\pi^*$; V) & 5.38 & 5.21 & 5.0 & 5.17 & 5.01 & 5.63 & 5.8 \\
\\
hydrogen sulfide & 1$^1$A$_2$(n$\rightarrow$4p; R) & 5.63 & 5.67 & 5.69 & 5.63 & 5.55 & 6.10 &  \\
\\
ketene & 1$^1$B$_1$($\pi\rightarrow$3s; R) & 5.87 & 5.97 & 6.06 & 5.97 & 6.11 & 6.06 & 5.86 \\
\\
methanimine & 1$^1$A"(n$\rightarrow\pi^*$; V) & 4.65 & 4.68 & 4.65 & 4.77 & 4.89 & 5.21 & \\
\\
thioformaldehyde & 1$^1$A$_2$(n$\rightarrow\pi^*$; V) & 1.91 & 1.77 & 1.61 & 1.74 & 1.57 & 2.20 & 2.03 \\
  & 1$^1$B$_2$(n$\rightarrow$4s; R) & 5.64 & 5.76 & 5.84 & 5.72  & 5.77 & 5.99 & 5.85 \\
  & 2$^1$A$_1$($\pi\rightarrow\pi^*$; V) & 5.36 & 6.07 & 6.77 & 6.02 &  6.66 &  6.34 & 6.2 \\
\\
water molecule& 1$^1$B$_1$(n$\rightarrow$3s; R) & 7.46 & 7.38 & 7.31 & 7.46 & 7.41 & 7.70 & 7.41 \\
  & 1$^1$A$_2$(n$\rightarrow$3p; R) & 8.91 & 8.89 & 8.88 & 9.02 & 9.11 & 9.47 & 9.2 \\
  & 2$^1$A$_1$(n$\rightarrow$3s; R) & 9.73 & 9.75 & 9.78 & 9.71 & 9.69 & 9.97 & 9.67 \\
\\
\hline
\rowcolor{Gray}
& ME (TBE) & -0.46 & -0.38 & -0.33 & -0.33 & -0.21 & &\\
& ME (EXP) & -0.30 & -0.22 & -0.17 & -0.17& -0.06 & &\\
\rowcolor{Gray}
& MAE (TBE) & 0.46 & 0.38 & 0.40 & 0.40& 0.44 & & \\
& MAE (EXP) & 0.33 & 0.32 & 0.39 & 0.27& 0.36 & &\\
\rowcolor{Gray}
& RMSE (TBE) & 0.54 & 0.43 & 0.45 & 0.43 & 0.51 & & \\
& RMSE (EXP) & 0.46 &-0.38 & 0.46 & 0.35 & 0.49 & & \\
\end{tabular}

  \textsuperscript{\emph{a}}Corrected theoretical best estimates as given in Ref.~\cite{Loos2018}. \ 
  \textsuperscript{\emph{b}}Experimental values listed in Ref.~\cite{Loos2018} (see references therein).
\end{table}

\newpage

\begin{sidewaysfigure}[th]
    \centering
    \includegraphics[width=1.0\textwidth]{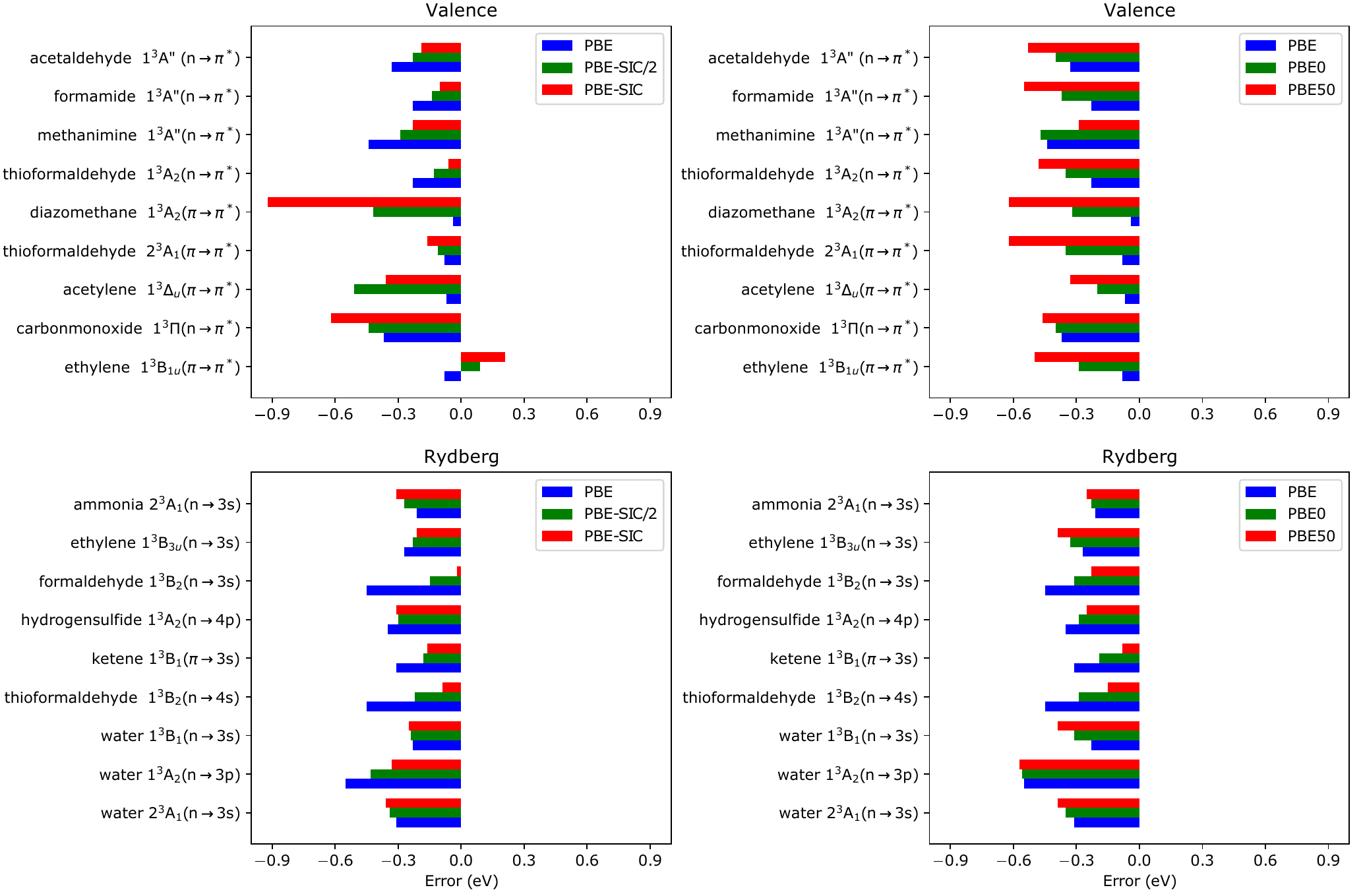}
    \caption{Error bar plot of Table S2}
    \label{fig: TripletsAll}
\end{sidewaysfigure}

\newpage

\begin{figure}[th]
    \centering
    \includegraphics[width=1.0\textwidth]{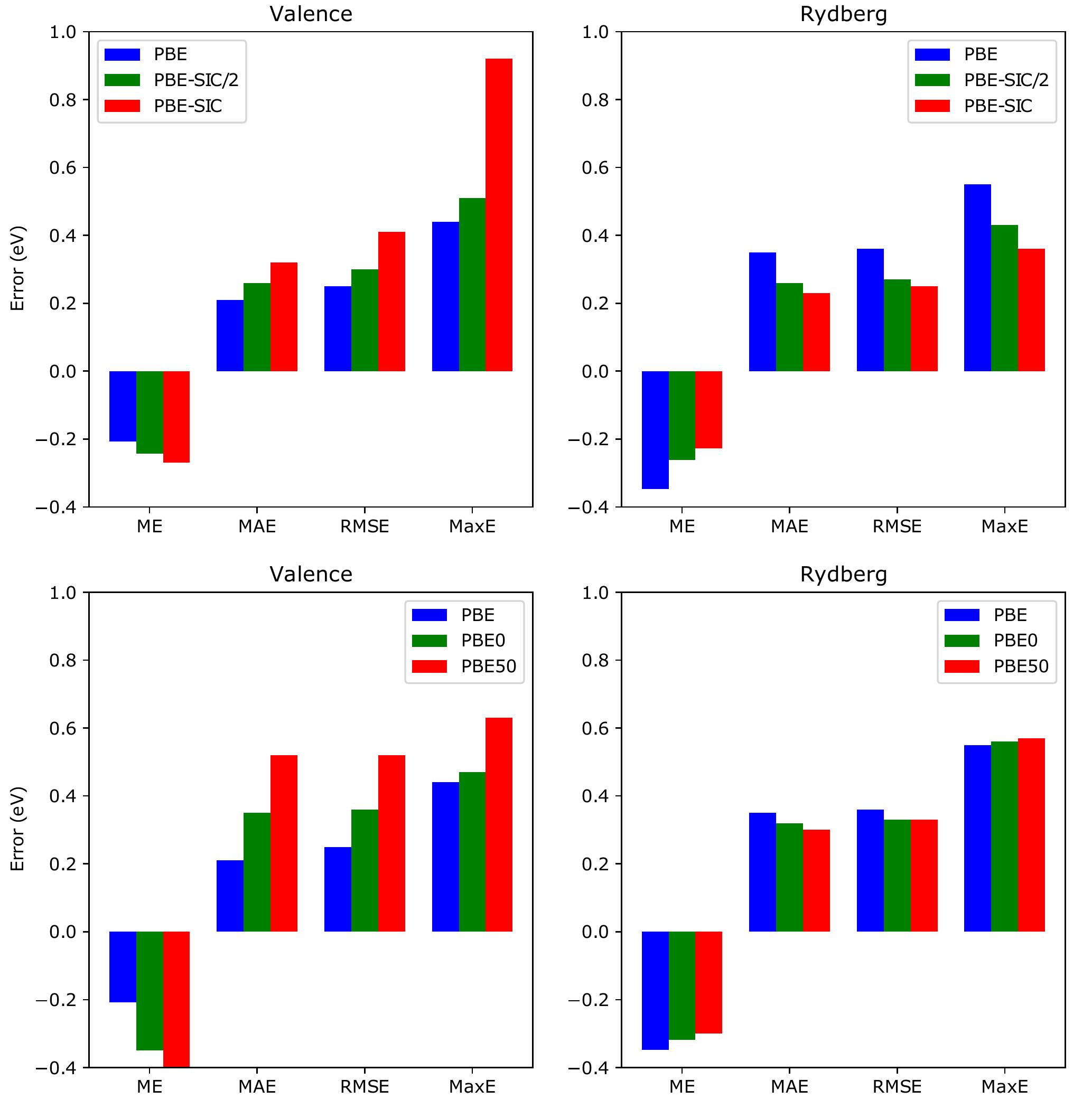}
    \caption{Error bar plot of Table S2}
    \label{fig: TripletsMean}
\end{figure}

\newpage

\begin{sidewaysfigure}[th]
    \centering
    \includegraphics[width=1.0\textwidth]{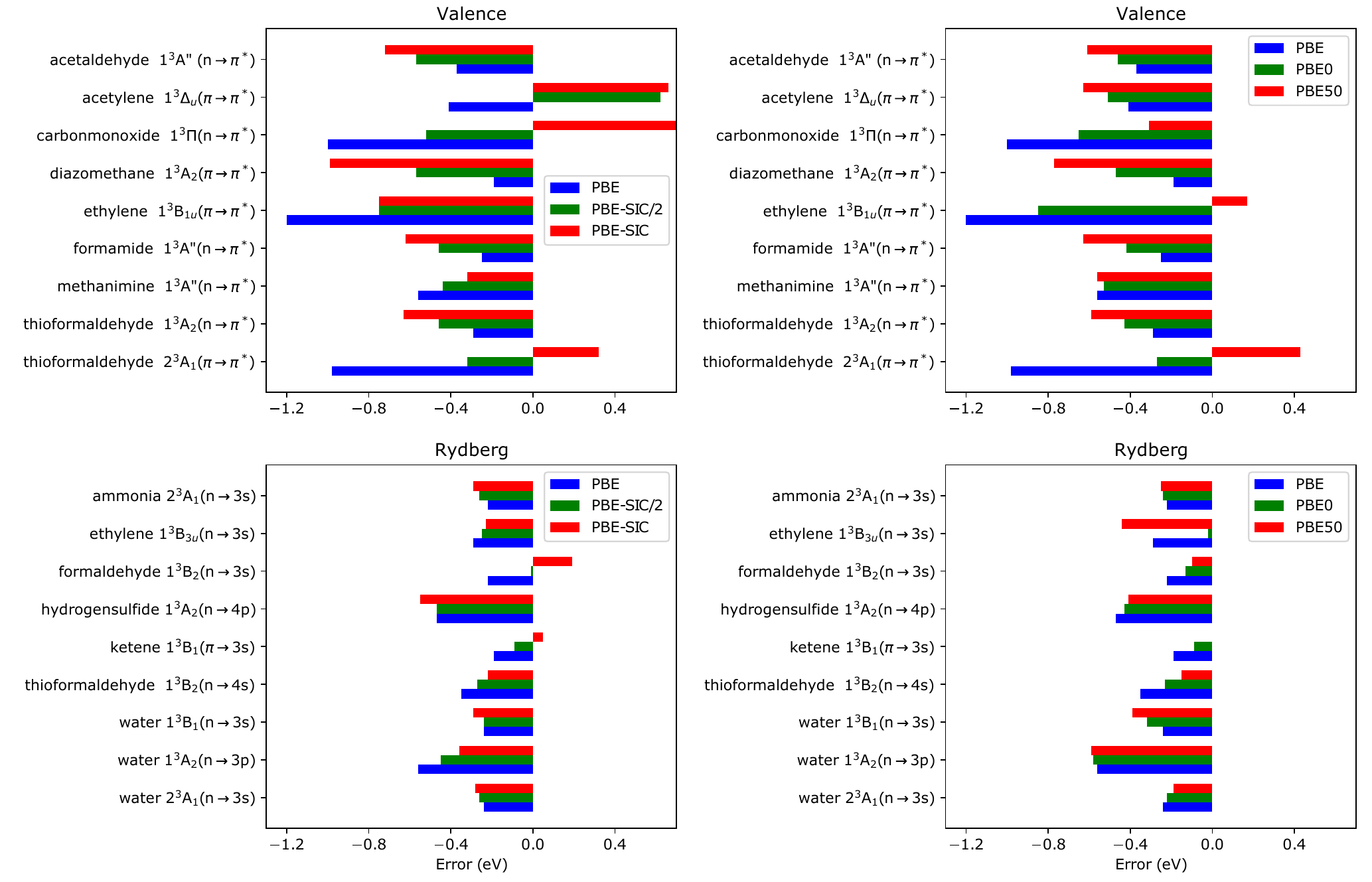}
    \caption{Error bar plot from Table S3} 
    \label{fig: SingletsAll}
\end{sidewaysfigure}

\newpage

\begin{figure}[th]
    \centering
    \includegraphics[width=1.0\textwidth]{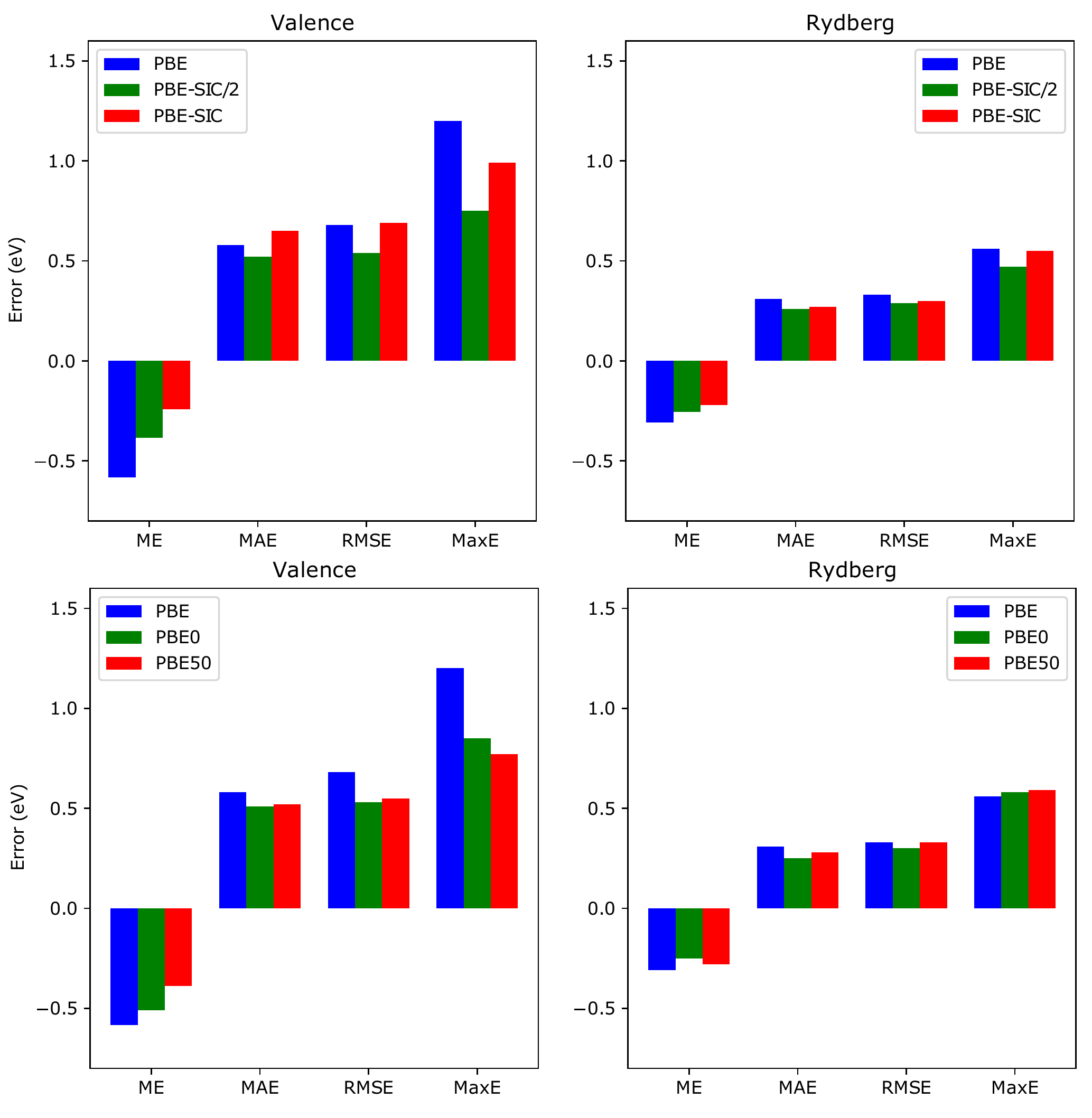}
    \caption{Error bar plot from Table S3} 
    \label{fig: SingletsMean}
\end{figure}
\clearpage

\newpage

\section{The Perdew Zunger Self-Interaction Correction Restricted to Real Orbitals (rSIC)}

\begin{table}[H]
\scriptsize
  \caption{Excitation energies evaluated with the PBE-rSIC/2 functional}
  \label{tbl: rSIC}
  \begin{tabular}{ l  c  c  c  c }
    \hline
    species & excitation & Mixed Spin & Triplet  &  Singlet\\
    \hline
acetaldehyde & $1^1$A$^{\prime\prime}$ ($n\rightarrow\pi^*$; V) & 3.71 & 3.73 & 3.69 \\
\\
acetylene & $1^1\Delta_u (\pi\rightarrow\pi^*$; V) & 6.56 & 5.60 & 7.52 \\
\\
ammonia & 2$^1$A$_1$(n$\rightarrow$3s; R) & 6.15 & 6.01 & 6.29 \\
\\
carbon monoxide & $1^1\Pi$(n$\rightarrow\pi^*$; V) & 6.94 & 5.79 & 8.09  \\
\\
diazomethane & 1$^1$A$_2$($\pi\rightarrow\pi^*$; V) & 2.26 & 2.15 & 2.36 \\
\\
ethylene & 1$^1$B$_{3u}$(n$\rightarrow$3s; R) & 6.87 & 6.80 & 6.93 \\
  & 1$^1$B$_{1u}$($\pi\rightarrow\pi^*$; V) & 5.90 & 4.62 & 7.19 \\
\\
formaldehyde & 1$^1$B$_2$(n$\rightarrow$3s; R) & 6.90 & 6.85 &  6.96 \\
\\
formamide & 1$^1$A"(n$\rightarrow\pi^*$; V) & 5.01 & 5.06 & 4.96 \\
\\
hydrogen sulfide & 1$^1$A$_2$(n$\rightarrow$4p; R) & 5.46 & 5.42 & 5.49 \\
\\
ketene & 1$^1$B$_1$($\pi\rightarrow$3s; R) & 5.73 & 5.24 & 6.22 \\
\\
methanimine & 1$^1$A"(n$\rightarrow\pi^*$; V) & 4.66 & 4.39 & 4.92 \\
\\
thioformaldehyde & 1$^1$A$_2$(n$\rightarrow\pi^*$; V) & 1.75 & 1.80 & 1.69 \\
  & 1$^1$B$_2$(n$\rightarrow$4s; R) & 5.60 &  5.51 & 5.69 \\
  & 2$^1$A$_1$($\pi\rightarrow\pi^*$; V) & 4.72 & 3.15 & 6.29 \\
\\
water & 1$^1$B$_1$(n$\rightarrow$3s; R) & 7.00 & 6.84 & 7.16 \\
  & 1$^1$A$_2$(n$\rightarrow$3p; R) & 8.69 & 8.63 & 8.76 \\
  & 2$^1$A$_1$(n$\rightarrow$3s; R)  & 9.43 & 9.21 & 9.66 \\
\\
\hline
\rowcolor{Gray}
& ME (TBE\textsuperscript{a}) & -0.73 &-0.38 & -0.40\\
& ME (EXP\textsuperscript{b}) & -0.59 &-0.23  & -0.24\\
\rowcolor{Gray}
& MAE (TBE\textsuperscript{a}) & 0.73  & 0.39 & 0.46\\
& MAE (EXP\textsuperscript{a}) & 0.59 & 0.26 & 0.34\\
\rowcolor{Gray}
& RMSE (TBE\textsuperscript{a}) & 0.95 & 0.44 & 0.50\\
& RMSE (EXP\textsuperscript{a}) & 0.88 & 0.31 & 0.41\\
\end{tabular}

  \textsuperscript{\emph{a}} Deviation from corrected theoretical best estimates as given in Ref.~\cite{Loos2018};
  \textsuperscript{\emph{b}} Deviation from experimental values collected in Ref.~\cite{Loos2018} (see references therein);

\end{table}

\clearpage

\section{Ground-State SIC Complex Orbitals of the Water Monomer}

\begin{figure}[H]
    \centering
    \includegraphics[width=1\textwidth]{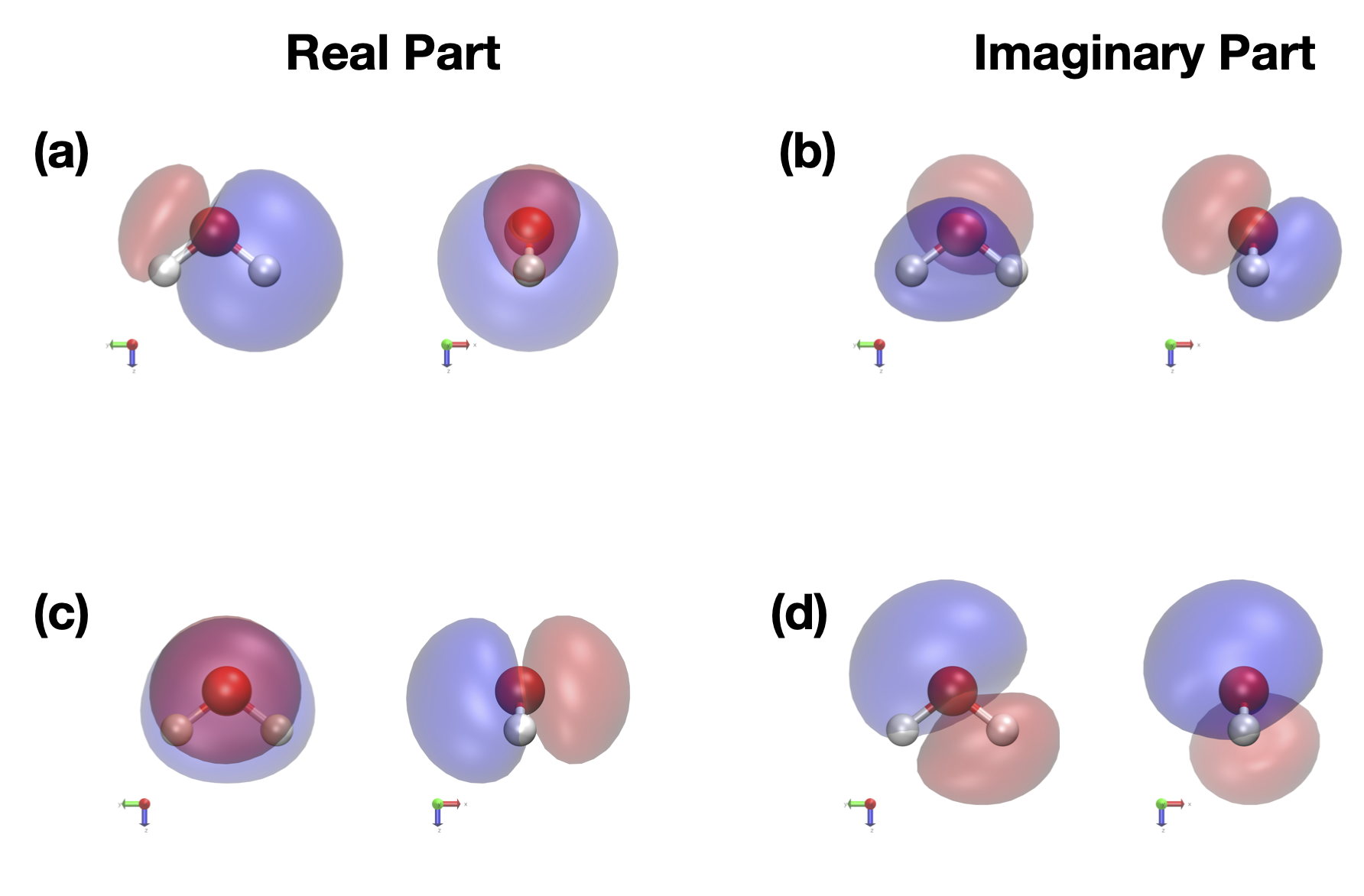}
    \caption{Real and imaginary parts of optimal ground-state SIC complex orbitals of the water monomer. The isofurfaces are represented with blue and red colors and correspond to isovalues of $\pm0.05$ {\AA}$^{-3/2}$. There are two degenerate pairs and only one orbital of each degenerate pair is shown. One optimal orbital (GS$_1$ as denoted in Table 3 of the main text) is made of real part (a) and imaginary part (b). The other optimal orbital (GS$_2$ as denoted in Table 3 of the main text)  is made of real part (c) and imaginary part (d). Each figure (a), (b), (c), and (d) contains its own isosurfaces from two different views as shown by the axis in insets. 
    } 
    \label{fig: water}
\end{figure}

\clearpage

\bibliography{supplement}